\documentclass[fleqn,usenatbib]{mnras}

\usepackage{newtxtext,newtxmath}
\usepackage{caption}
\usepackage{subcaption}
\usepackage{floatrow}   
\usepackage{mathtools}
\usepackage{cancel}
\usepackage{booktabs}
\usepackage{subcaption}
\usepackage{amsmath}
\usepackage{ulem}
\usepackage{multirow}
\usepackage[T1]{fontenc}
\usepackage{svg}
\newcommand{\orcid}[1]{\href{https://orcid.org/#1}{\includesvg[width=10pt]{orcid}}}

\DeclareRobustCommand{\VAN}[3]{#2}
\let\VANthebibliography\thebibliography
\def\thebibliography{\DeclareRobustCommand{\VAN}[3]{##3}\VANthebibliography}
\usepackage{cancel}

\usepackage{graphicx}	
\usepackage{amsmath}
\usepackage{soul}
\usepackage{multicol}
\usepackage{orcidlink}

\newcommand{\kms}{\,km\,s$^{-1}$}

\title[M67 with TESS and K2]{Photometric variability of blue straggler stars in M67 with \textit{TESS} and \textit{K2}}

\author[Vernekar et al.]{
Nagaraj Vernekar$^{1,2}$\thanks{E-mail: nagarajvernekar30@gmail.com} \orcidlink{0000-0003-2947-5889},
Annapurni Subramaniam$^{1}$ \orcidlink{0000-0003-4612-620X},
Vikrant V. Jadhav$^{1,3,4}$\orcidlink{0000-0002-8672-3300},
Dominic M. Bowman$^{5}$ \orcidlink{0000-0001-7402-3852} \\
$^{1}$ Indian Institute of Astrophysics, Koramangala II Block, Bangalore-560034, India.\\
$^{2}$ Dipartimento di Fisica e Astronomia "Galileo Galilei", Università di Padova, Vicolo dell’Osservatorio 3, IT-35122, Padova, Italy. \\
$^{3}$ Inter-University Centre for Astronomy and Astrophysics (IUCAA), Post Bag 4, Ganeshkhind, Pune 411007, India.\\
$^{4}$ Helmholtz-Institut für Strahlen- und Kernphysik, Universität Bonn, Nussallee 14-16, D-53115 Bonn, Germany.\\
$^{5}$ Institute of Astronomy, KU Leuven, Celestijnenlaan 200D, 3001 Leuven, Belgium.
}\usepackage{cancel}
\date{Accepted 2023 June 26. Received 2023 June 26; in original form 2022 November 13 }

\pubyear{2022}

\begin{document}
\label{firstpage}
\pagerange{\pageref{firstpage}--\pageref{lastpage}}
\maketitle

\begin{abstract}
Blue straggler stars (BSSs) are formed through mass transfer or mergers in binaries. The recent detections of white dwarf (WD) companions to BSSs in M67 suggested a mass transfer pathway of formation. In search of a close companion to five BSSs in M67 that are known to be spectroscopic binaries, we study the light curves from \textit{K2} and \textit{TESS} data. We use PHOEBE to analyse the light curves and estimate the properties of the companions. 
We detect variability in WOCS 1007, and the light curve is dominated by ellipsoidal variation. Using the light curve and radial velocity measurements, we estimate its orbital period to be 4.212$\pm$0.041~d and $e$ = 0.206$\pm$002. The mass of the companion is estimated to be  0.22$\pm$0.05~M$_{\sun}$ with a radius of  0.078$\pm$0.027~R$_{\sun}$, confirming it to be an LM WD with T$_{\rm eff}$ = 14300$\pm$1100 K. The estimated mass of the BSS, 1.95$\pm$0.26~M$_{\sun}$, is similar to that estimated from isochrones. The BSS in WOCS 1007 shows $\delta$ Scuti pulsations, although it is slightly deformed and likely to be formed through an efficient mass transfer. Though we detect a light curve for WOCS 4003 showing grazing eclipse with ellipsoidal variation, the estimated parameters are inconclusive. Apart from the 0.44 d period, we found smaller eclipses with a period of 1.1 d, suggesting a compact triple system. In the case of  WOCS 4003, WOCS 5005, and WOCS 1025, no eclipses or pulsations are detected, confirming the absence of any short-period inner binary with high inclination in these BSSs.

\end{abstract}




\section{Introduction}

Open clusters are known to host a large fraction of binaries. The evolution of single and binary stars is dictated by the complex interaction of stars within the cluster. Blue straggler stars (BSSs) are dwarfs that are brighter and bluer than the stars at the main sequence turn-off of a cluster and are thought to have gained mass to continue to stay on the main sequence. The formation pathways of BSSs are through merger or mass transfer in binaries or triples \citep{McCrea1964, Perets2009, Naoz2014}. The rich, old and dynamically active open cluster, M67, is known to host several BSSs \citep{Manteiga1989, Gilliland1992, Deng99, Bruntt2007, Lu2010, Geller2015}. In M67, about $\sim$79~per~cent of the BSSs are in binaries with a range of orbital periods and eccentricities \citep{Latham1996, Sandquist2003}. The high binary fraction combined with a large period range suggests that BSSs could be formed through different mechanisms. 

The mass transfer in binaries results in the formation of a stellar remnant, mostly white dwarfs (WD), as companions to the BSSs. Mergers of stars resulting in forming a BSS do not have a binary companion. As the stellar density in open clusters is low for direct collisions, such a formation pathway is less common \citep{Jadhav2021MNRAS.507.1699J}. Detections of WD companions to BSSs have implied mass transfer as a formation pathway \citep{Gosnell2015, Gosnell2019, Sindhu2019, Sindhu2020}. The BSSs are thought to be formed via case A, B, or C mass transfer, based on the phase of mass transfer within the main sequence, before the core-helium burning stage or after core-helium exhaustion of the donor \citep{Kippenhahn1967, Lauterborn1970}. Case A mass transfer could lead to the coalescence of the binary, resulting in a single massive BSS or a binary with less massive BSS and a short-period companion. Case B mass transfer produces a BSS in a short period system with a helium core remnant of mass $<$ 0.45~M$_{\odot}$ from an initial mass of $<$ 3~M$_{\odot}$. In Case C, we expect a CO WD, with mass $\ge$ 0.5~M$_{\odot}$. BSSs in long-period binaries are expected to have undergone Case C mass transfer \citep{Perets2015}.

UV investigations have revealed the presence of hot WD companions to the BSSs and main sequence stars \citep{Knigge2008, Gosnell2015, Sindhu2019, Leiner2019, Jadhav2019, sahu_2019, Sindhu2020, Subramaniam2020}. Studies by \citet{Sindhu2019} and \citet{2021MNRAS.507.2373P} on BSSs in M67 revealed many such hot companions. Based on the number of low-mass (LM) WD companion detections, \citet{Sindhu2020} suggested that 35 \% of the stars in their sample must have undergone Case B mass transfer.
The issues that require reconciling are as follows: (1) BSSs with LM WD should be in short-period systems such as WOCS 1007, whereas the spectroscopic orbital periods are found to be $\ge$ 1000 d (WOCS 5005 and WOCS 1025); (2) photometric masses of BSSs (that are estimated using isochrones) are thought to be over-estimated when compared to kinematic masses (which are difficult to estimate) (3) parameter estimations of the BSSs and the hot companions have been done by modelling multi-wavelength SEDs, and it is good to have an independent estimation.
 
This paper sheds light on the above using the light curve analysis of \textit{K2} and \textit{TESS} data for a few select BSSs in M67. In this study, we present (1) the detection of pulsation and (2) the detection and modelling of light curves to study the properties of the BSSs and the binary system. The paper is arranged as follows: Sec. \ref{sec:data} gives a log of data and analytical methods, Sec. \ref{sec:results} contains the results for individual systems, and we discuss the results in Sec. \ref{sec:discussion}. 
 
\begin{table*}
    \centering
    \caption{Details of known parameters of the targets. Remarks are taken from $^{a}$\citet{Geller2015} and $^{b}$\citet{2013A&A...556A..52T}.}
    \begin{tabular}{cccc cccc}
    \hline
    Name & \textit{Gaia} DR3 source id & RA & Dec & P (d) & $e$ & Corrected data & Remarks\\ \hline
    
    WOCS 1007 & 604918213472162432 & 08 51 34.31 & $+$11 51 10.49 & 4.182 & 0.205 & \textit{K2}, \textit{TESS} & SB1$^{a}$, $\delta$~Scuti$^{b}$\\
    WOCS 4006 & 604918179110923520 & 08 51 32.57 & +11 50 40.63 &- &- & \textit{K2}, \textit{TESS} & SB1$^{a}$, $\delta$~Scuti$^{b}$ \\
    WOCS 4003 & 604918041671889792 & 08 51 28.14 & +11 49 27.49 & 0.441 & 0.0 & \textit{K2} & SB1$^{a}$ \\
    WOCS 5005 & 604917285757663872 & 08 51 19.90 & +11 47 00.44 & 4913 & 0.34 & \textit{K2} & SB1$^{a}$ \\
    WOCS 1025 & 604896566835438464 & 08 51 37.69 & +11 37 03.79 & 1154 & 0.07 & \textit{K2} & SB1$^{a}$\\ \hline
    \end{tabular}
    \label{tab:target_data}
\end{table*}

\section{Data and methods}
\label{sec:data}
\subsection{\textit{K2}}
This study uses data from the \textit{K2} mission. The \textit{K2} mission was divided into 19 campaigns that were about 80 days in length \citep{2014PASP..126..398H}. \textit{K2} has two types of data, short-cadence (1 min) and long-cadence (30 min). The accuracy of the \textit{K2} photometry is not as high as \textit{Kepler} photometry due to inaccurate pointing, leading to a drift in the field of view. To correct for the drift, thrusters aboard the satellite were fired once every 6 hrs \citep{2014PASP..126..398H}. This process introduces systematics into the light curves, which are not inherent to the stars. Therefore, we use the light curves corrected by the \textit{K2} Systematics Correction (K2SC) algorithm \citep{2016MNRAS.459.2408A}.
K2SC is a detrending algorithm that uses Gaussian processes to determine and remove all the systematics due to the pointing jitter. \textit{K2} observed M67 in campaigns 05, 16, and 18. 

\subsection{\textit{TESS}}
The Transiting Exoplanet Survey Satellite (\textit{TESS}) \citep{2015JATIS...1a4003R} is a broad-band photometric survey mission covering most of the sky. Each hemisphere is divided into 13 sectors and is observed for a year, with each sector being observed for 27.4 days. \textit{TESS} provides 30-min cadence data for over 10 million stars through its full-frame images. And for a pre-determined sample of 200,000 stars \citep{2016MNRAS.463.4210N,2017MNRAS.472.4173R}, \textit{TESS} collects 2-min cadence data.
In the extended mission that started in July 2020, the cadence was augmented to 20 s and 10 min for the short and long cadence, respectively. The light curves for the stars with short cadence data are extracted by the \textit{TESS} Science Processing Operation Center (SPOC) pipeline \citep{2016SPIE.9913E..3EJ} through Simple Aperture Photometry (SAP). As SAP flux retains systematics, the pipeline performs corrections to it, referred to as Pre-search Data Conditioned SAP (PDCSAP) flux. During its extended mission, \textit{TESS} has observed M67 in sectors 44, 45, and 46 (2021 October 12 to 2021 December 30). 

\subsection{Sample selection}
M67 is an excellent cluster to study the BSS population due to the extensive literature available about its 14 known BSSs \citep{Latham1996, 2009A&A...503..165Y, Geller2015, Jadhav2019, Sindhu2020}. Among these, our primary focus is on BSSs with possible binary companions, where light curves could be used to constrain the binary properties. Five systems listed in Table \ref{tab:target_data} are found to be suitable for carrying out a detailed study. All the systems are observed by both \textit{TESS} and \textit{K2} missions, but not all systems have corrected data from these missions. Two systems, WOCS 1007 and WOCS 4006, have light curves corrected by the K2SC algorithm and the SPOC pipeline. The other three systems do not have corrected \textit{TESS} light curves but have a K2SC light curve. We note that the crowding of the systems in M67 increases the complexity of extracting and correcting the \textit{TESS} data obtained through full-frame images, as it could suffer from severe contamination that adversely affects the analysis. 

\subsection{Light curve analysis}
If a BSS is found in a binary system with a compact object like a WD, the eclipses in the light curves are expected to be shallow and may not be visually evident. In most systems, the amplitude of variability in the out-of-eclipse regions is comparable to the depth of the eclipses. An amplitude spectrum was calculated using Fourier transform to extract the eclipses from such light curves. This is because when a light curve of an eclipsing system is phase folded using the significant frequencies obtained through the amplitude spectrum, the eclipses line up on top of each other for the correct orbital period, which appears as two dips (primary and secondary eclipses) in the phase folded curve. The phase-folded light curve can also show out-of-eclipse sinusoidal variability. This could be caused either due to pulsations in the star, surface activity such as spots or ellipsoidal variability (see Sec. \ref{section:EV}). 

\subsubsection{Orbital period estimation}
A discrete Fourier transform of the light curves provides the amplitude spectrum of the variability. The significant frequencies were extracted from the amplitude spectrum through an iterative pre-whitening process using the software tool Pythia\footnote{\url{https://github.com/colej/pythia}}. For each step, the frequency with the highest S/N in the amplitude spectrum was selected with the corresponding amplitude and phase calculated from non-linear least-squares fitting of a sinusoid to the light curve (see~\cite{2021A&A...656A.158B}). Frequencies were extracted until the S/N was $<4$ \citep{1993A&A...271..482B}. 
Once all the significant frequencies were obtained, Period04 \citep{2014ascl.soft07009L} was used to phase-fold the light curve for each of those frequencies to identify possible eclipses.

\subsubsection{Binary modelling}
To understand the formation pathway of a BSS, it is essential to obtain constraints on the nature of its companion. As eclipsing systems provide a direct way of determining physical properties, the companion's nature can be constrained through binary modelling. This study used an eclipsing binary modelling software called PHOEBE 1.0\footnote{\url{http://phoebe-project.org/}} \citep{2005ApJ...628..426P}. PHOEBE uses a model based on the Wilson-Devinney code \citep{1971ApJ...166..605W}. It simultaneously fits both the light curve and the radial velocity curve to determine the stellar and orbital parameters of a system.

The modelling of systems was carried out in steps. The first step was to manually fit the radial velocity curve of the system to estimate the mass ratio, eccentricity, and argument of the periastron. After obtaining a close enough solution, an iterative minimisation method was carried out to find the best-fit model. This was carried out by repeatedly performing the differential correction method (as described in \citealt{2005ApJ...628..426P}), where the input parameter set was updated for the next iteration until the convergence of the cost function. The light curve was introduced into the modelling once the best-fit radial velocity model was obtained. Similar to the radial velocity curve, the two curves are manually fitted, followed by an iterative minimisation. The light curve was directly modelled for a system without radial velocity data.

\subsubsection{Ellipsoidal Variability}\label{section:EV}

In close binary systems, stars can be tidally distorted due to the gravitational force of each other. The non-spherical shape of the companions causes a variation in the light curve known as ellipsoidal variability \citep{1985Ap&SS.117...69B,1985ApJ...295..143M}. Ellipsoidal variables have lower inclination angles, such that the dominant effect on the light curve is due to the distorted shape of the star. The light curves of ellipsoidal variables are usually quasi-sinusoidal in nature and can be modelled using the Fourier series \citep{1985Ap&SS.117...69B,1985ApJ...295..143M,1993ApJ...419..344M,2013PASA...30...16D}: 

\begin{align}
\centering
    L(\phi) = A_{0} + \sum_{n=1}^{z}  A_{n} \, \cos(n\phi) + \sum_{n=1}^{z} B_{n} \, \sin(n\phi)
    \label{eqn:EV}
\end{align}

\noindent where $\phi$ is the orbital phase. The coefficients of the cos (i$\phi$) provide information about the nature of variation present in the light curve. The coefficient of cos(2$\phi$) term provides the amplitude of the ellipsoidal variation. In contrast, the coefficient of cos($\phi$) term provides the amplitude of light variations taking place within the orbital period, such as temperature spots and magnetic activity \citep{1999IBVS.4684....1H,2013PASA...30...16D}. If ellipsoidal variability is dominant in a light curve, then the coefficient of cos(2$\phi$) term is the largest. Here, non-linear least-squares fitting of the Fourier series was used to characterise the system's ellipsoidal variability.  

\section{Results}
\label{sec:results}

\begin{figure*}
\centering
    \begin{subfigure}{.84\linewidth}
    \centering
    \includegraphics[width=\linewidth]{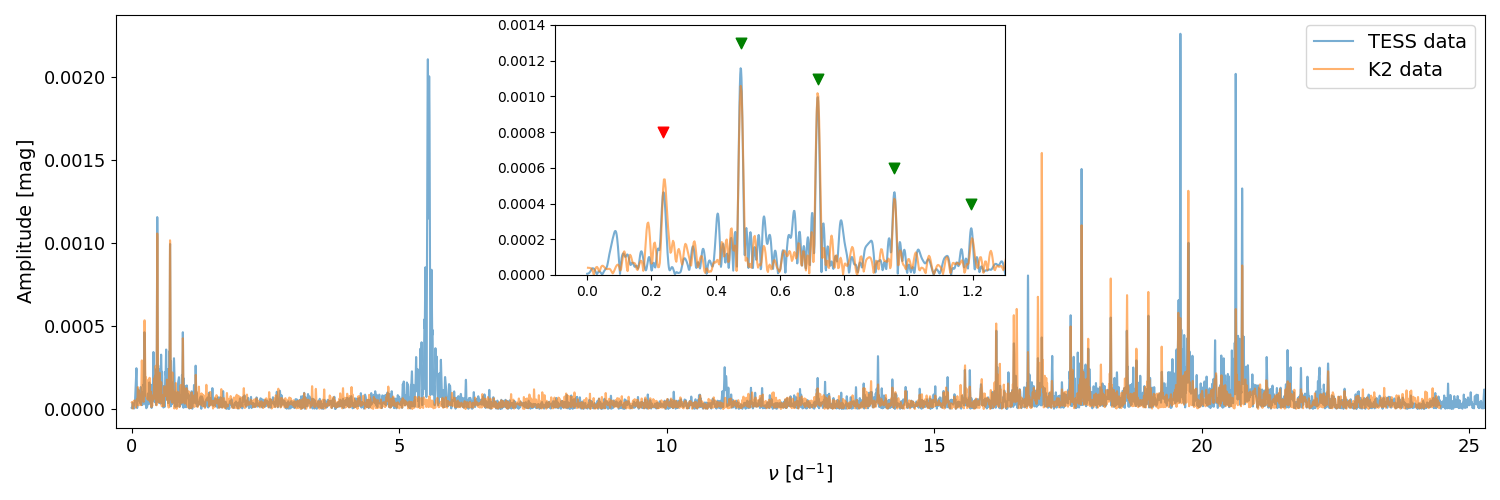}
    \caption{}\label{fig:1007ampli}
    \end{subfigure} %

    \hfill
    
    \begin{subfigure}{.9\linewidth}
    \centering
    \includegraphics[width=\linewidth]{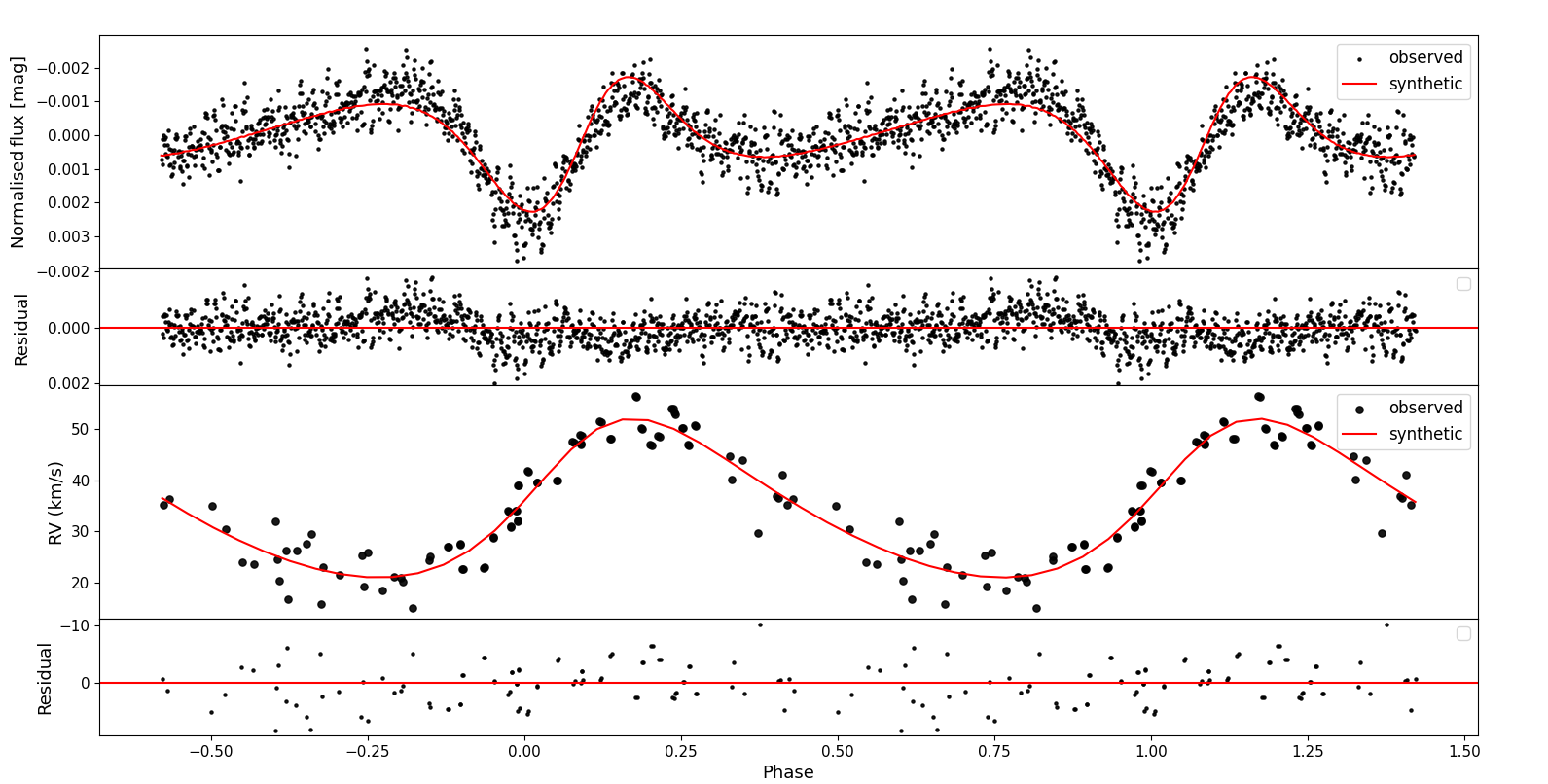}
    \caption{}\label{fig:1007fitting}
    \end{subfigure} %
    \RawCaption{\caption{\textit{\ref{fig:1007ampli}: }Amplitude spectrum of WOCS 1007 obtained through Fourier analysis of the \textit{TESS} (blue) and \textit{K2} (orange) data. The low-frequency region of WOCS 1007 is shown in the inserted panel with the red triangle representing the orbital frequency of the system (0.237 d$^{-1}$ or 4.212 d) and the green triangles representing the harmonics of the orbital frequency. \textit{\ref{fig:1007fitting}: } Phase folded curve (top panels) and radial velocity curve (bottom panels) fitting of WOCS 1007 obtained from the PHOEBE modelling along with the residuals.}
\label{fig:imasyy}}
\end{figure*}

\subsection{WOCS 1007}
WOCS 1007 is a short-period eccentric ($e$ = 0.205) binary system \citep{1992ASPC...32..152L} with an orbital period of 4.18 d \citep{1992IAUS..151..475M}. The primary companion has a spectral type of B9V \citep{1995AJ....109.1379A} with a mass of 2.0--2.2 M$_{\odot}$ \citep{1992ASPC...32..152L}. The star is known to be a $\delta$ Scuti pulsator \citep{2013A&A...556A..52T} with a rotational velocity of v\,$\sin$\,i = 79 km\,s$^{-1}$ \citep{2018MNRAS.480.4314B}. \citet{Sindhu2019} used SED analysis to estimate the parameters of the BSS and the companion. By fitting a double SED, they estimated the BSS to have T$_{\rm eff}$ = 7500--7750~K with a radius of 2.94--2.97~R$_{\odot}$ and the hot companion to have   T$_{\rm eff}= 13\,250-13\,750$~K with a radius of 0.095$\pm$0.021~R$_{\odot}$. By comparing the parameters of the hot companion with the models, they classified the companion to be an LM WD star and estimated a mass of 0.19~M$_{\odot}$.  \textit{Gaia} DR3 \citep{GaiaDR3_2022} classified this source as a non-variable and non-eclipsing SB1 binary. The parameters from \textit{Gaia} DR3 are included in Table~\ref{1007table}. As the secondary star is optically sub-luminous, we can assume the photometric astrophysical parameters provided in \textit{Gaia} DR3 represent the primary (with the caveats presented in the \textit{Gaia} validation papers).

We used the PDCSAP flux from \textit{TESS} and K2SC light curve from \textit{K2} for the frequency analysis. We performed frequency analysis on both the light curves to cross-check whether the significant frequencies were present across the instruments. Fig. \ref{fig:1007ampli} shows the amplitude spectrum of WOCS 1007 obtained using the \textit{TESS} and \textit{K2} data in blue and orange, respectively. From the \textit{TESS} data, a total of 42 significant frequencies were obtained with the dominant frequency at 19.60 d$^{-1}$, whereas for the \textit{K2} data, only 25 significant frequencies were obtained with the dominant frequency at 17.01~d$^{-1}$. From Fig. \ref{fig:1007ampli}, it is evident that the majority of frequencies are present in both the data sets with different amplitudes except for a few frequencies. The prominent peak at 5.5d$^{-1}$ seen in the \textit{TESS} data is not present in the \textit{K2} data. Upon checking the light curves and amplitude spectra of all the neighbouring stars, this peak was found to be the second harmonic of the orbital frequency of a nearby source (about 53''). As the peak and its harmonics were present due to contamination, they were excluded from the analysis.

Here, we are more interested in the low-frequency peaks (f < 5 d$^{-1}$), which are present both in \textit{K2} and \textit{TESS} data with similar amplitudes. For further analysis, only the \textit{TESS} data were used because of its shorter 2-min cadence compared to the 30-min \textit{K2} cadence. In the low-frequency region of the amplitude spectrum, there are five significant frequencies at 0.236, 0.477, 0.717, 0.955 and 1.195 d$^{-1}$ as shown in Fig. \ref{fig:1007ampli}. The first frequency is the orbital frequency (f$_{\rm orb}$), and the other four are harmonics of the orbital frequency (nf$_{\rm orb}$). A dip resembling an eclipse was obtained by phase folding the light curve for 0.237 d$^{-1}$ frequency or 4.212 d period. This orbital period is consistent with the period found by \citet{1992IAUS..151..475M}. As all the low frequencies are associated with the binarity, we estimate the period of the WOCS 1007 binary system from its light curve for the first time, along with the $\delta$ Scuti pulsations \citep{breger2000,bowman2018}.

For the binary modelling of WOCS 1007, we performed PHOEBE analysis using three different models (detached binary, semi-detached with the primary filling its Roche lobe, and unconstrained binary). The fitting was performed multiple times using each model with different initial guesses for the free parameters. Along with the radial velocity data from \citet{1991ASPC...13..424M}, we used the binned phase folded curve (binned into 500 bins) to reduce the amplitude of the pulsations and the computation time of the modelling. In all the runs, the period of the system was fixed as 4.212~d, obtained from the frequency analysis. The albedo and gravity brightening of the primary was fixed to be 1 due to the radiative envelope of the primary star. Given that the light curve is available only in one passband, the T$_{\rm eff}$ of the primary was fixed to 7600 K \citep{Sindhu2019}. As the semi-detached model could not properly reproduce the phase folded curve, leading to a large cost function, it was discarded from the analysis. All the runs of the other two models converged close together, irrespective of the initial conditions. Due to this convergence, we classified the system as a detached binary. Fig. \ref{fig:1007fitting} shows the best-fitting model obtained for the light curve and radial velocity data. A critical feature of the fit is that the ingress and egress of the eclipse in the synthetic curve are not equal. Table \ref{1007table} shows the values and errors of the fitting parameters. The errors from PHOEBE GUI are usually underestimated.
Hence, we used a Python script\footnote{\url{https://sourceforge.net/p/phoebe/mailman/message/33650955/}} to estimate errors \citep{2005ApJ...628..426P}. This code allows a user to run Markov Chain Monte Carlo (MCMC) method (on PHOEBE 1.0 backend) using the EMCEE package \citep{emcee}, which is a pure-Python implementation of Goodman \& Weare’s Affine Invariant MCMC Ensemble sampler \citep{goodman}. For the MCMC sampling, we provided the following nine free parameters: semi-major axis, centre-of-mass velocity, inclination, eccentricity, mass ratio, argument of periastron, two surface potentials and secondary T$_{\rm eff}$. The number of walkers was set to 50, and the total number of iterations to 3000. The first 200 iterations were discarded as the burn-in period. Figure \ref{fig:mcmc} shows the corner plots of the free parameters. The standard deviation of the marginalised posterior parameter distributions, i.e. 1$\sigma$, was taken as the error.
The orbital parameters and masses of the companions obtained from PHOEBE agree with the solution from \citet{1992IAUS..151..475M}. The secondary temperature of 14300$\pm$1100~K and radius of 0.078$\pm$0.027~R$_{\odot}$ are consistent with the values obtained through SED fitting by \citet{Sindhu2019}. If we vary the T$_{\rm eff}$ of the primary by 500 K, the stellar parameters do not show a significant change, but orbital parameters such as the inclination and mass ratio were found to show a change of the order of 5--8 per cent. The light curve analysis, in combination with the radial velocity, finds the secondary to have a mass of 0.22$\pm$0.05 M$_{\odot}$, suggesting it to be an LM WD. This is in excellent agreement with the results from the SED analysis. We, therefore, confirm that the WOCS 1007 is indeed a BSS+LM WD system. The log(g) of the secondary is found to be 6.00$\pm$0.27 \,cm\,s$^{-2}$, which agrees with the values expected for LM WDs. The mass of the primary is estimated as 1.95$\pm$0.26 M$_{\sun}$, which is also similar to the literature values and estimated mass using isochrones \citep{Sindhu2018}.

\begin{table*}
        \centering
         \caption{Orbital and stellar parameters of WOCS 1007 obtained from PHOEBE modelling. The SED and literature parameters are taken from $^{a}$\citet{Sindhu2019} and $^{b}$ \citet{1992IAUS..151..475M}, respectively. }
         \begin{tabular}{ccccc}
         \toprule
    Parameters & This work & SED$^a$ &Literature$^b$ & \textit{Gaia} DR3\\ \hline
    \vspace{0.2cm}
    Semi-major axis $(R_{\sun})$ & $14.22^{{+0.19}}_{{-0.16}}$ & & &\\
    \vspace{0.2cm}
    Eccentricity & ${0.206}^{{+0.002}}_{{-0.002}}$ &  & 0.256 & 0.16$\pm$0.10\\
    \vspace{0.2cm}
    Argument of periastron (rad) & $5.48^{{+0.01}}_{{-0.01}}$& & & 5.63$\pm$0.60\\
    \vspace{0.2cm}
    Center of mass velocity (\kms) & ${34.12}^{{+0.14}}_{{-0.13}}$& & & 31.93$\pm$1.31\\
    \vspace{0.2cm}
        Inclination (deg) & ${73.6}^{{+2.4}}_{{-2.3}}$ & & &\\
    \vspace{0.2cm}
    Mass ratio & 0.115$^{{+0.002}}_{{-0.002}}$& & $>$ 0.092\\
    \vspace{0.2cm}
    Period (d) & 4.212$\pm$0.041 & & 4.182 & 4.185$\pm$0.001\\
    \vspace{0.2cm}
    Mass 1 (M$_{\sun}$) & 1.95$^{{+0.26}}_{{-0.24}}$ &  & 2.08 & 1.95$\pm$0.06\\
    \vspace{0.2cm}
    Mass 2 (M$_{\sun}$) & 0.22$^{{+0.05}}_{{-0.05}}$ & $\sim$0.2 & $<$ 0.2 & $>$0.18\\
    \vspace{0.2cm}
    Radius 1 (R$_{\odot}$) &  ${2.54}^{{+0.81}}_{{-0.81}}$ &2.94$\pm$0.04 & &\\
    \vspace{0.2cm}
    Radius 2 (R$_{\odot}$) & {0.078}$^{{+0.023}}_{{-0.027}}$ &0.094$\pm$0.001 &  &\\
    \vspace{0.2cm}
    Eff. Temperature 1 (K) & 7600 &  & & 7536$\pm$38, 7655$\pm$10\\
    \vspace{0.2cm}
    Eff. Temperature 2 (K) &  {14300}$^{{+1100}}_{{-1000}}$ & 13250--13750 & &\\
    \vspace{0.2cm}
    log g$_{1}$ (\,cm\,s$^{-2}$) & {3.91}$^{{+0.16}}_{{-0.15}}$ &$\sim$ 3.5 & & 3.61$\pm$0.01\\
    \vspace{0.2cm}
    log g$_{2}$ (\,cm\,s$^{-2}$) &{6.00}$^{{+0.22}}_{{-0.27}}$& $\sim$ 7.75--9.0 & &\\ \hline
    \end{tabular}
    \label{1007table}
\end{table*}

\begin{figure*}
\centering
    \centering
     \includegraphics[width=0.9\textwidth]{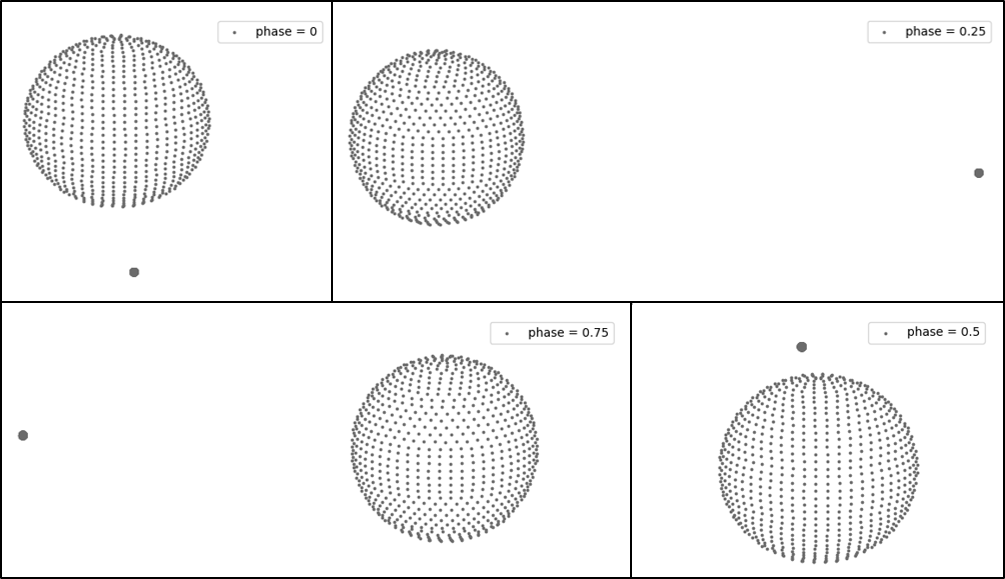}
   \caption{Geometric configuration of WOCS 1007 at four phases (0,0.25,0.5 and 0.75), constructed using the PHOEBE results.}\label{fig:1007geo}
\end{figure*}

\begin{figure}   
    \includegraphics[width=1\linewidth]{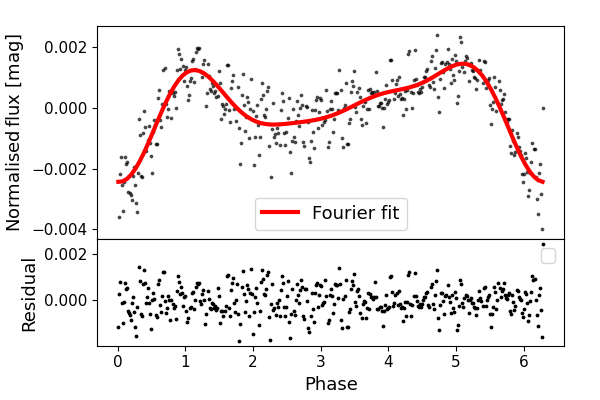}
   \caption{The Fourier fit obtained for WOCS 1007 using Eqn.~(\ref{eqn:EV}) in the top panel, with the residuals in the bottom panel.}\label{fig:1007fourier}

\end{figure}

As seen from the geometry of the binary system, shown in Fig. \ref{fig:1007geo}, it is evident that the components do not eclipse each other. An inclination of 73.6$\pm$2.4 is not high enough to produce eclipses in binaries containing compact objects. Therefore, the light curve variation is likely due to the ellipsoidal variability caused by the distorted shape of either or both stars. The fit obtained using the Fourier series given in Eqn.~\ref{eqn:EV} is shown in Fig. \ref{fig:1007fourier} and the coefficients are given in Table \ref{1007fouriertable}. The coefficient of cos(2$\phi$) term (A$_{2}$) is the largest, indicating that the ellipsoidal variability dominates the light curve. Even though ellipsoidal variability is dominant, the presence of the $\delta$ Scuti pulsations in the BSS indicates that the distortions in shape are not large enough to inhibit pulsations. This is supported by Fig. \ref{fig:1007geo}, where we do not see a significant distortion. 

Therefore, WOCS1007 is a binary system that shows ellipsoidal variability, where the primary is a slightly distorted BSS showing $\delta$ Scuti pulsations, and the secondary is an LM WD in a short period orbit. 

\begin{table}
        \centering
         \caption{Coefficients of Eqn.~(\ref{eqn:EV}) obtained through Fourier modelling of WOCS 1007 with the errors given in brackets. A$_{n}$ are coefficients of cos(n\,$\phi$) and B$_{n}$ are coefficients of sin(n\,$\phi$).}
         \begin{tabular}{cccc}
         \toprule
         Coefficients & values & Coefficients & values\\
         \hline
    A$_{0}$ & 0.00011(1) & &\\
    A$_{1}$ & 0.00031(4) & B$_{1}$ & 0.00005(2)\\ 
    A$_{2}$ & 0.00105(4) & B$_{2}$ & 0.00022(1)\\
    A$_{3}$ & 0.00077(4) & B$_{3}$ & 0.00004(1)\\
    A$_{4}$ & 0.00029(5) & B$_{4}$ & 0.00005(2)\\
    \hline
    \end{tabular}
    \label{1007fouriertable}
\end{table}

\subsection{WOCS 4003}
WOCS 4003 was identified as an F3V type star by \citet{1940ApJ....91..244E} and as an eclipsing binary by \citet{2015A&A...573A.100F}. \citet{2009A&A...503..165Y} conducted a detailed photometric analysis of WOCS 4003 and found an orbital period of 0.44 d. Even though the nature of the light curve suggested a contact binary, some unusual features were reported. They observed the O'Connell effect in the light curve where the maxima were not of equal brightness. \cite{1993PASP..105.1433R} and \cite{2004A&A...416.1097S} compared the Fourier coefficients (a$_{4}$ and a$_{2}$) to differentiate between contact and non-contact systems. They found WOCS 4003 to be deviating from the boundary condition expected for a W Uma system. Later, the system was classified as semi-detached using the diagram from \citet{2006MNRAS.368.1311P}. From PHOEBE modelling, \citet{2009A&A...503..165Y} found the effective temperatures of the component stars to be 6900 K and 5200--5830~K with radii of 0.38--0.45~R$_{\odot}$ and 0.29--0.33~R$_{\odot}$, respectively. Their modelling also included three spots (two cold spots and one hot spot) on the cooler star. \citet{Jadhav2019} noted a large UV excess flux in the SED of WOCS 4003, which is not expected for the binary parameters obtained by \citet{2009A&A...503..165Y}. 
Though \citet{Jadhav2019} estimated the probable parameters of a binary companion, they suggested the UV excess to be of a different origin, related to the presence of X-ray emission from the system. Their estimates of only the primary are likely to be reliable (T$_{\rm eff}$ = 6500$\pm$125~K and radius = 1.78$\pm$0.02~R$_{\sun}$) unless the secondary has a large contribution to flux. 
\citet{GaiaDR3_2022} classified WOCS 4003 as a variable eclipsing binary with a period of 0.441 d. Unfortunately, \textit{Gaia} DR3 does not provide additional orbital parameters, and the spectro-photometric parameters may be incorrect as both components are optically bright. 

\begin{figure*}
\centering
    \begin{subfigure}{.9\linewidth}
    \centering
    \includegraphics[width=\linewidth]{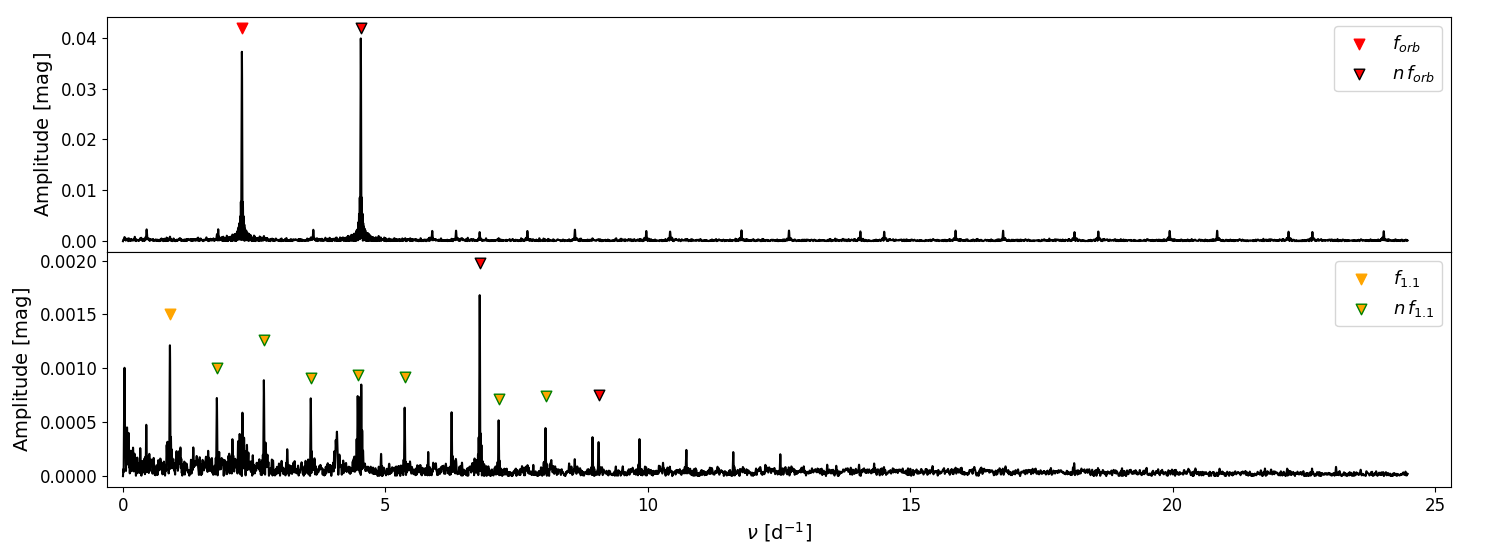}
    \caption{}\label{fig:4003ampli}
    \end{subfigure} %

    \hfill
    
    \begin{subfigure}{.9\linewidth}
    \centering
    \includegraphics[width=\linewidth]{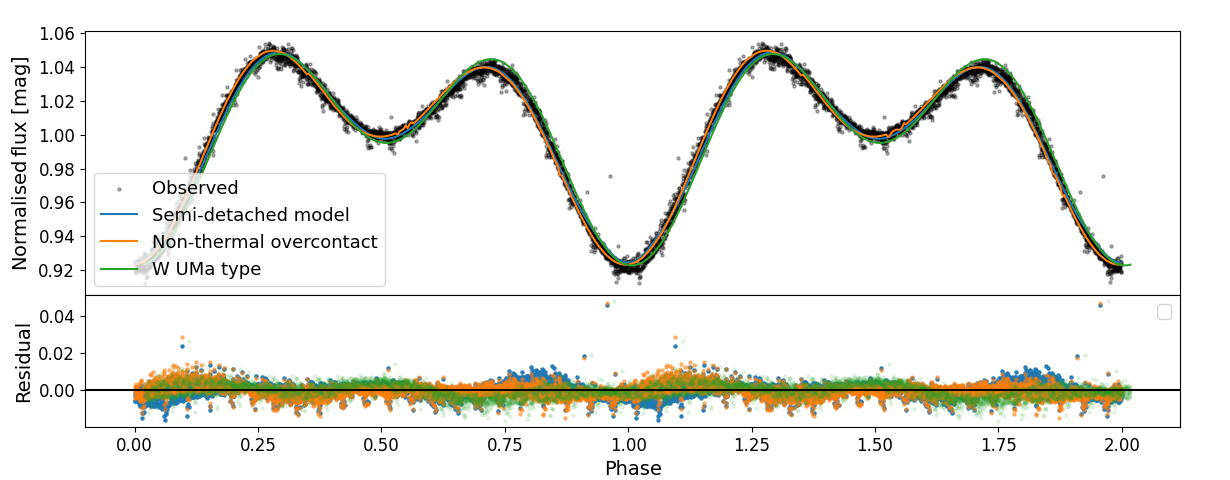}
    \caption{}\label{fig:4003LC} 
    \end{subfigure} %

    \hfill

    \begin{subfigure}{.8\linewidth}
    \includegraphics[width=\linewidth]{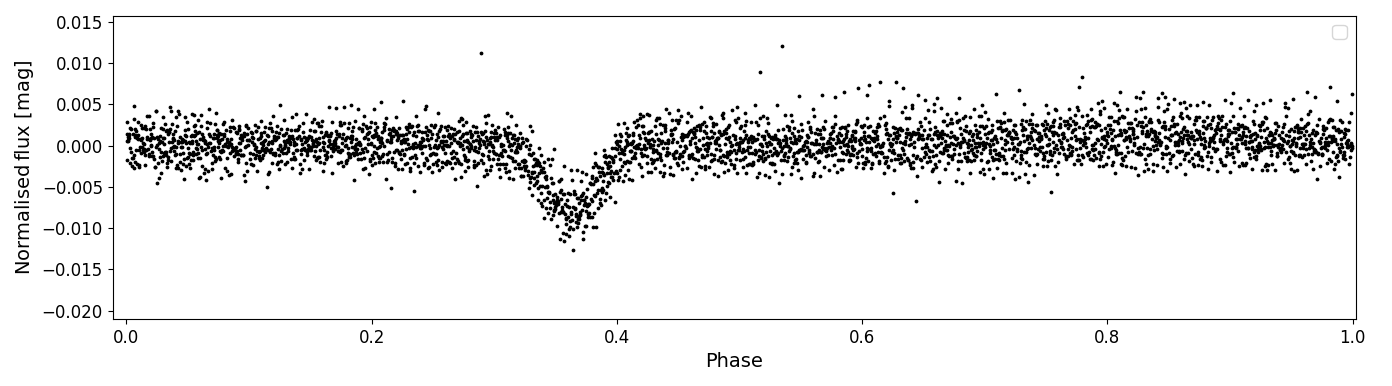}
    \caption{}\label{fig:4003dip}
    \end{subfigure}

    \RawCaption{\caption{\textit{\ref{fig:4003ampli}: }\textit{Top panel: }Amplitude spectrum of WOCS 4003 obtained using the \textit{K2} data. The red triangles represent the orbital frequency of the system (2.265 d$^{-1}$ or period of 0.44 d) and its harmonics. \textit{Bottom panel: }Amplitude spectrum of WOCS 4003 after removing the two dominant frequencies. The frequency associated with the period of the smaller dip (1.1 d), along with its harmonics, are represented by yellow triangles. \textit{\ref{fig:4003LC}: }Fitting of the phase folded curve along with the residual of WOCS 4003 assuming two different models. \textit{\ref{fig:4003dip}: }WOCS 4003 light curve obtained by removing the eclipses and phase folding using a period of 1.11 d.}
\label{fig:imas}}
\end{figure*}

Only the K2SC light curve was used for the analysis due to the unavailability of PDCSAP flux from \textit{TESS}, and 17 significant frequencies were found in the amplitude spectrum (Fig. \ref{fig:4003ampli}). There are two peaks in the low-frequency region, as shown by the green and red triangles. The 2.265 d$^{-1}$ or 0.44 d is taken as the orbital frequency and 4.53 d$^{-1}$ as its harmonic. The orbital period derived is consistent with the period found by \citet{2009A&A...503..165Y}. We also report the presence of the O'Connell effect in the light curve as seen in Fig. \ref{fig:4003LC}.
 
Apart from the large eclipses in the phase folded curve, smaller dips were also present at regular intervals throughout the orbital phase (Fig. \ref{fig:4003LC}). By removing the dominant eclipses through subtraction of the sine waves corresponding to the three frequencies (4.53, 2.26, and 6.79 d$^{-1}$) and their respective amplitudes (also known as pre-whitening), we found the period of the small dips to be 1.11 d. Fig. \ref{fig:4003dip} shows the pre-whitened phase folded curve.

For the PHOEBE analysis, due to the uncertainties in the system's configuration, the fitting was performed with three different models (W UMa type over-contact system, non-thermal over-contact system, and semi-detached with the primary filling its Roche lobe). We fixed the period (0.44 d), gravity brightening and albedo of unity for all three models. The surface temperature of the primary was allowed to vary in the W UMa model but was fixed to 6500 K \citep{Jadhav2019} in the other two. To improve the fit of the synthetic to the observed light curve, we introduced two spots (one hot and one cold spot) in the non-thermal model and one cold spot in the semi-detached model. The fits obtained for the three models are shown in Fig \ref{fig:4003LC}. Most of the parameters obtained from the three models disagreed with each other, and none of the three agreed with the SED results, especially the secondary radius. The three models returned secondary radii significantly larger than the SED value, leading to an over-luminous system.

\begin{figure}
    \centering
    \includegraphics[width=1\textwidth]{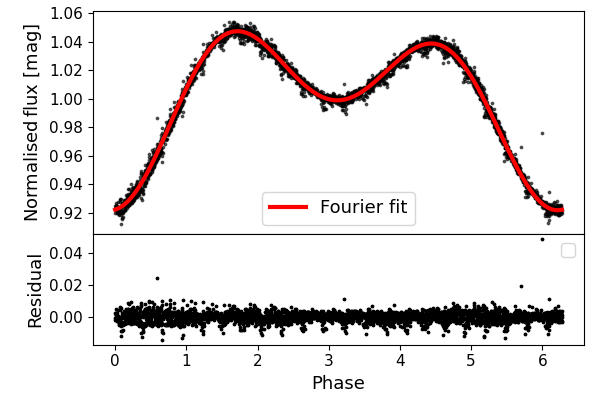}
   \caption{The Fourier fit obtained for WOCS 4003 using Eqn.~(\ref{eqn:EV}) is in the top panel, with the residuals to the fit shown in the bottom panel.}
   \label{fig:4003fourier}

\end{figure}

In all three models, the inclination of the system was determined to be low. However, due to the larger radii of the components and the small distance between them, they might be exhibiting a grazing eclipse. The geometric configuration (in semi-detached mode) showed that only a small portion of the flux is blocked during eclipses. Therefore, confirming if ellipsoidal variability has a dominant effect on the light curve is vital. For this, Eqn.~(\ref{eqn:EV}) was fitted to the light curve. Fig. \ref{fig:4003fourier} shows the fit, and Table \ref{4003fouriertable} gives the values of the coefficients. We obtained similar values for A$_{1}$ and A$_{2}$, which are the coefficients of cos\,($\phi$) and cos\,(2$\phi$). This indicates almost equal contributions from the eclipse and the ellipsoidal variability. 

The potential eclipsing and ellipsoidal variable WOCS 4003 is a complicated system, as the total luminosity from the SED and light curve analyses do not match. Notably, the radii and T$_{\rm eff}$ obtained from the light curve analysis suggest the system to be very luminous, whereas the SED analysis does not support such a total luminosity. Most importantly, the detected period of 1.11 d suggests a third component, which needs confirmation through further studies.

\begin{table}
        \centering
         \caption{Coefficients obtained through Fourier modelling of WOCS 4003. A$_{n}$ are coefficients of cos(n\,$\phi$) and B$_{n}$ are coefficients of sin(n\,$\phi$).}
         \begin{tabular}{cccc}
         \toprule
         Coefficients & values & Coefficients & values\\
         \hline
    A$_{0}$ & 1.00001(1) & &\\
    A$_{1}$ & 0.03686(7) & B$_{1}$ & 0.00535(5)\\ 
    A$_{2}$ & 0.03968(8) & B$_{2}$ & 0.00244(1)\\
    A$_{3}$ & 0.00165(4) & B$_{3}$ & 0.00007(1)\\
    \hline
    \end{tabular}
    \label{4003fouriertable}
\end{table}

\subsection{WOCS 4006}
\citet{Manteiga1989} classified WOCS 4006 as an SB1 system with a v\,$\sin$\,i = 80--100 km\,s$^{-1}$ \citep{Pritchet1991,2000A&AS..144..469R}. \citet{1991AJ....101..541G} discovered $\delta$ Scuti pulsations in the system and \citet{Bruntt2007} detected 41 significant frequencies. Two studies reported an effective temperature of 7244 K \citep{1991AJ....101..541G} and 8090 K \citep{1991A&A...245..467M} for the BSS. \citet{2021MNRAS.507.2373P} used double SED fitting technique and found effective temperatures of 7500$\pm$125 and 16000$\pm$125 K with radii of 1.61$\pm$0.02 and 0.051$\pm$0.005 R$_{\odot}$ for the primary and secondary, respectively. The \textit{Gaia} DR3 light curve did not exhibit significant variability, and the \textit{Gaia} DR3 parameters are as follows: T$_{\rm eff}$ = 7958$\pm$44~K, log (g) = 4.29$\pm$0.02~dex, R = 1.52$\pm$0.05~R$_{\odot}$ and v\,$\sin$\,i = 112$\pm$8 km\,s$^{-1}$. This is a rapidly rotating BSS.

\begin{figure*}    
    \centering
    \includegraphics[width=0.96\textwidth]{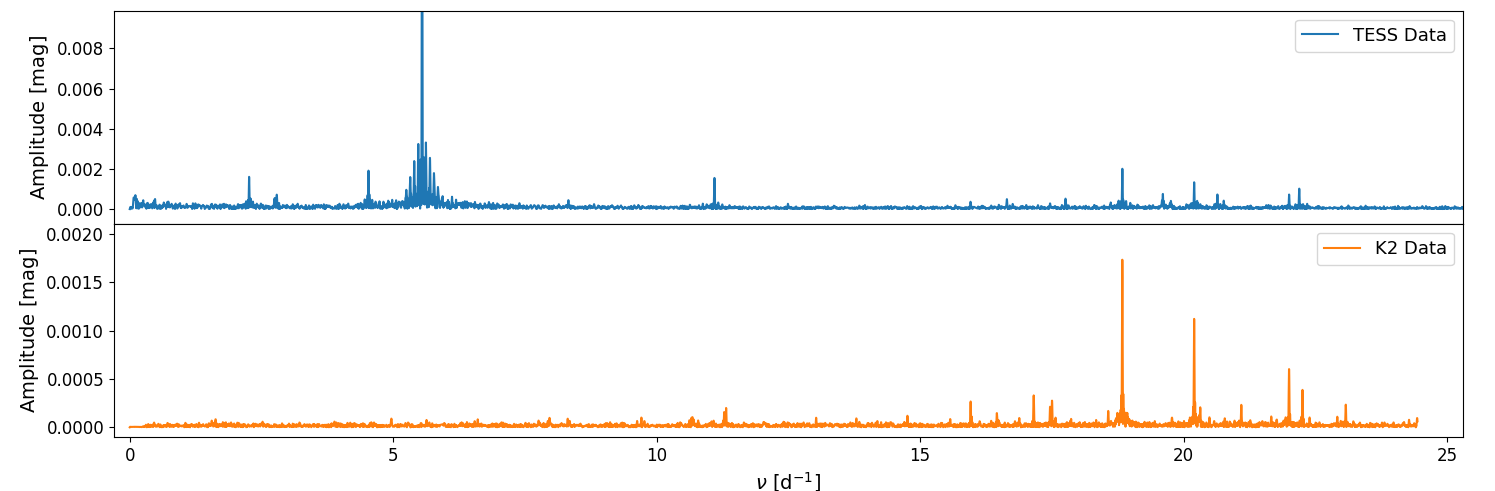}
   \caption{Amplitude spectrum of WOCS 4006 obtained through Fourier analysis of the \textit{TESS} (top panel) and \textit{K2} (bottom panel) data.}
   \label{fig:4006period}

\end{figure*}

As the PDCSAP flux from \textit{TESS} and K2SC data from \textit{K2} were available, both were used for the frequency analysis.
Fig. \ref{fig:4006period} shows the amplitude spectra of both data sets. Being located in a dense field, WOCS 4006 suffers from contamination by multiple sources. We see from Fig. \ref{fig:4006period} that frequencies between 2 and 6 d$^{-1}$ are present only in \textit{TESS} data (top panel) but not in the \textit{K2} data (bottom panel). Due to this large contamination,  \textit{TESS} data were not used for the analysis. The amplitude spectrum using K2 data shows $\delta$ Scuti pulsations between 15 and 25 d$^{-1}$.  
We did not detect any eclipses or ellipsoidal variations in the phase-folded K2 data. This study, therefore, confirms the source as a $\delta$ Scuti pulsator and rules out the presence of a close companion (period of a few days) with a large inclination in this SB1 system. 

\begin{figure*}
\centering
    \includegraphics[width=0.9\linewidth]{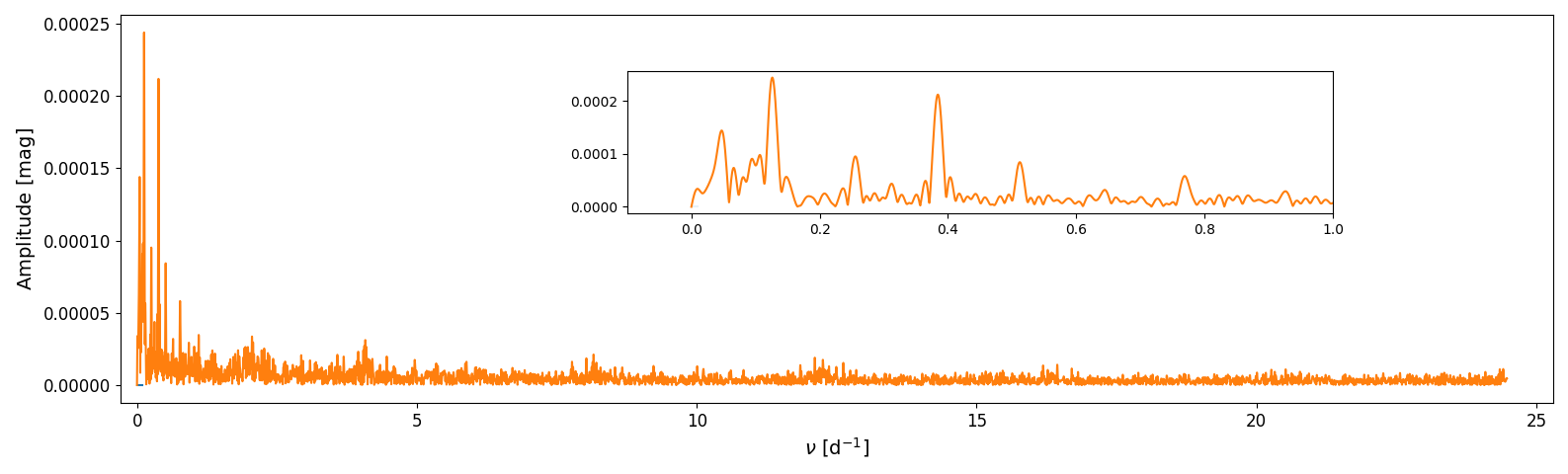}
   \includegraphics[width=0.9\linewidth]{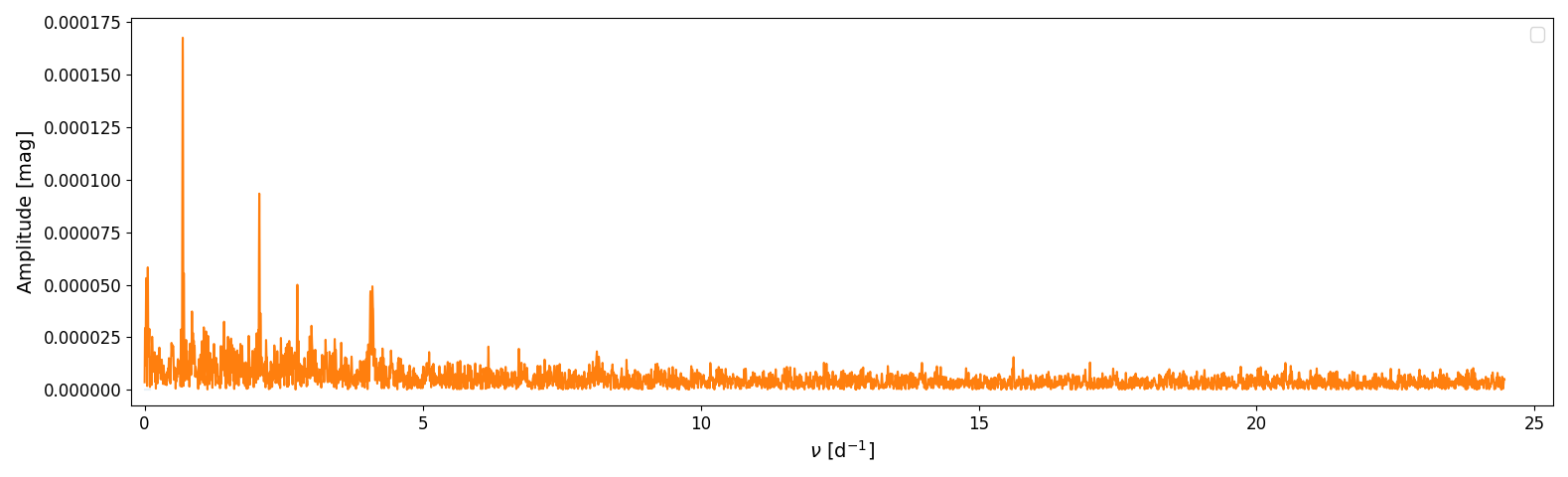}
    \caption{Amplitude spectrum of WOCS 5005 (top) and WOCS 1025 (bottom) obtained using the \textit{K2} data.}\label{fig:1025ampli}
\end{figure*}

\subsection{WOCS 5005}

WOCS 5005 is an SB1 system \citep{Geller2015} with a spectral type of F5IV \citep{1995AJ....109.1379A}. \citet{Latham1996} used spectroscopic data covering over 4000 d to determine the orbital period to be 4913 d with an eccentricity of 0.342. The star is a slow rotator with a rotational velocity of less than 30 \kms \citep{Pritchet1991}. \citet{VandenBerg2004Chandra} detected X-rays from Chandra observations and suggested the presence of a close binary. \citet{2021MNRAS.507.2373P} used double SED fitting to estimate surface temperatures of 6500 K and 13000 K for the primary and secondary, respectively. They also found the radius of the secondary to be 0.035 R$_{\sun}$ and suggested it to be an LM WD, which places a demand on it being in a short period with the BSS. WOCS 5005 also falls outside the region resulting from stable mass transfer and requires a period of $<$ 10 d to comply with formation through mass transfer from the estimated WD companion \citep{2021MNRAS.507.2373P}. \textit{Gaia} DR3 did not demonstrate any variability or binarity in the system. Assuming the optical spectrum is dominated by the primary, the \textit{Gaia} DR3 parameters for the primary are as follows: T$_{\rm eff}$ = 6890$\pm$200~K, log(g) = 4.17$\pm$0.20~dex and R = 2.36$\pm$0.06~R$_{\sun}$.

WOCS 5005 has light curves from three algorithms (Kepler SOC pipeline \citep{2010ApJ...713L..87J}, K2SFF \citep{2014PASP..126..948V}, and K2SC \citep{2016MNRAS.459.2408A}). A total of nine significant frequencies were obtained from the \textit{K2} pipeline (top panel of Fig.~\ref{fig:1025ampli}). The light curve folded for 7.95 d showed shallow dips.
Due to their small depth, the dips could not be directly classified as eclipses. We analysed K2SC and K2SFF data to verify the period and the dips. However, the two data sets did not contain the 7.95 d period. Even after analysing the extracted \textit{TESS} data of WOCS 5005, the period of 7.95 d was not obtained. We discard this estimation as it is likely to be spurious. This analysis, therefore, does not support the presence of a close companion (P $\sim$ a few days) with a large inclination in the system. 
 
\subsection{WOCS 1025}
\citet{Geller2015} classified WOCS 1025 as an SB1 system. The system's orbital period was determined to be 1154 d with an eccentricity of 0.066$\pm$0.082 \citep{Latham1996}. \citet{Latham1996} also found the rotational velocity to be v\,$\sin$\,i = 60 km\,s$^{-1}$. \citet{2021MNRAS.507.2373P} used SED fitting and determined the effective temperature of the primary to be 7000$\pm$125 K. They could not estimate the secondary effective temperature precisely but stated that it could be between 21\,000 K and 50\,000 K. \textit{Gaia} DR3 classified the source as a non-variable binary. The binary parameters are as follows: centre of mass velocity = 33.77$\pm$1.87~km\,s$^{-1}$, $e$ = 0.52$\pm$0.06, P = 773$\pm$9 d, M$_1$ = 1.54$\pm$0.14~M$_{\odot}$ and M$_2$ = 0.61 to 1.72~M$_{\odot}$. If we assume the primary dominates the optical spectrum, the primary parameters are as follows: log(g) = 4.11$\pm$0.02~dex, R = 1.61$\pm$0.05~R$_{\odot}$, T$_{\rm eff}$ = 6954$\pm$600 K.

WOCS 1025 is observed by both \textit{K2} and \textit{TESS} missions, but due to the unavailability of PDCSAP flux from \textit{TESS}, only the K2SC light curve was used. From the frequency analysis, six significant frequencies were obtained (bottom panel of Fig.~\ref{fig:1025ampli}). However, no eclipse-like dips were obtained by phase folding the light curve. This analysis does not support the presence of a close companion (P $\sim$ a few days) with a large inclination in the system.

\section{Discussion} \label{sec:discussion}

This study utilised the capabilities of \textit{TESS} and \textit{K2} to trace small amplitude variability among the BSSs in M67. The targets of interest are BSSs with detected LM WD companions and known $\delta$ Scuti pulsators. A binary system with an LM WD companion demands a short orbit binary system, and the light curves of such a system may show eclipses or ellipsoidal variability if the orbit is highly inclined. 

In the case of WOCS 1007, we could combine the light curve with the radial velocity data to estimate the parameters of the binary components. We consider the orbital results quite robust, with better estimates of the period and eccentricity. The estimated radius and T$_{\rm eff}$ of the companion match well with those estimated from the SED analysis. The mass of the hot companion demands it to be an LM WD, and the log(g) value also supports this. 

BSSs can be significantly more massive than the turn-off mass, possibly requiring a large amount of mass accumulated onto them from an external source \citep{Perets2015}.
It is suggested that the mass of the BSSs, as estimated from the colour-magnitude diagram (CMD), is about 15 per cent more than that estimated from dynamical estimates \citep{Perets2015}. Therefore, estimating the BSSs parameters through methods other than CMD is essential. From the binary modelling of WOCS 1007, we estimated a mass of 1.95$\pm$0.26 M$_\odot$ (see table \ref{1007table}) for the BSS, which is not very different from that estimated from isochrones ($\sim$ 2.0 M$_\odot$) \citep{Sindhu2018}.

The BSS is likely to have gained mass through a Case-B mass-transfer in its binary system if we consider the lower limit of the mass of the BSS mass, which is at least $\sim$ 0.3 M$_\odot$ more than the turn-off mass of the cluster. As the LM WD companion needs to shed the outer layers early in the RGB evolution, the BSS could have gained that mass, making WOCS 1007 a post-mass-transfer system. We also note that a conventional mass-transfer scenario cannot explain the mass gained if the upper limit of the primary mass is considered, as the mass gained ($\sim$ 1.0 M$_\odot$) is much larger than the conventional limit and therefore demands a more efficient mass transfer mechanism. We conclude that this binary system is a post-mass-transfer system. 

We suggest that the progenitor of WOCS 1007 is likely to be an equal mass ($\sim$ 1.35 M$_\odot$) binary. The evolution of the current secondary towards the red-giant branch resulted in the filling up of the Roche lobe and subsequent mass transfer. The current primary gained $\sim$  0.6 M$_\odot$, with the secondary ending up as LM Helium core WD of mass $\sim$ 0.22 M$_\odot$.

\begin{figure*}    
    \centering
    \includegraphics[width=0.96\textwidth]{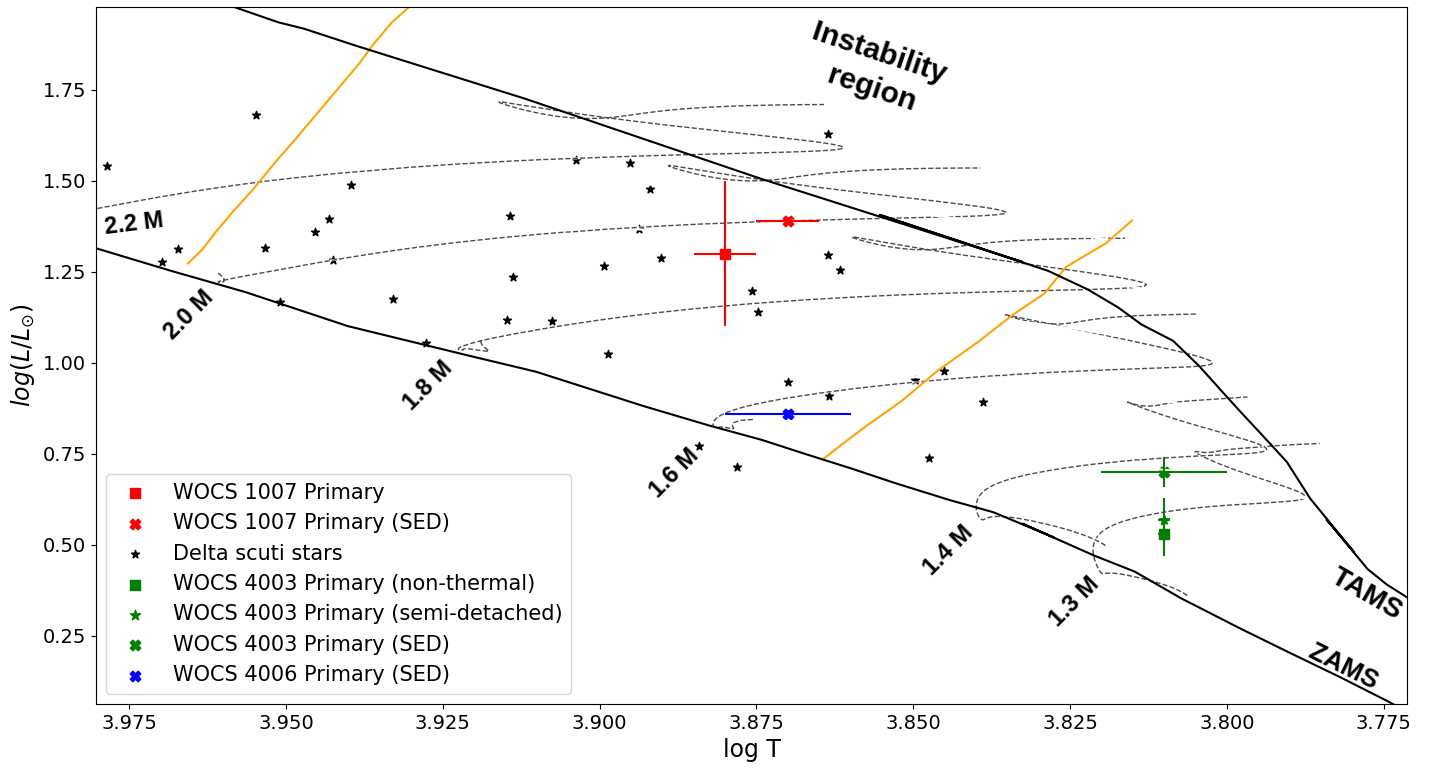}
   \caption{HR diagram with the $\delta$ Scuti instability region taken from \citet{murphy2019}. SED results of WOCS 1007, WOCS 4006, and WOCS 4003 are represented using red, blue, and green crosses, respectively. 
   The PHOEBE result of WOCS 1007 is represented using a red square, whereas a square and a star of the colour green represent the PHOEBE results of WOCS 4003 obtained using the non-thermal and semi-detached model, respectively. 
   The grey dashed lines represent the MIST single-star evolutionary stars for different masses. The black stars represent $\delta$ Scuti stars taken from \citet{2017MNRAS.465.1181L}.}
   \label{fig:delta}
\end{figure*}

We also note that the BSS in WOCS 1007 is a $\delta$ Scuti star. The pulsations are in addition to the ellipsoidal variability due to the slightly deformed BSS. We detect 19 frequencies in the amplitude spectrum. We do not detect any signatures of tidally perturbed pulsations \citep{2019ApJ...883L..26B} or single-sided pulsations seen in systems such as \citet{2020NatAs...4..684H}. To understand the evolutionary stage of the BSS, WOCS 1007 was placed on the HR diagram as shown in Figure \ref{fig:delta}, along with other known $\delta$ Scuti stars in binary systems using the catalogue from \cite{2017MNRAS.465.1181L}. The $\delta$ Scuti instability region indicated by the two yellow lines was taken from \citet{murphy2019}. The BSS is located on the main sequence and the empirical instability region. This agrees with the $\delta$ Scuti nature of the BSS and that the star is placed in the instability strip due to its rejuvenation of increased mass through mass transfer. The location of the BSS also agrees with the theoretical single-star evolutionary tracks obtained from MIST (MESA Isochrones and Stellar Tracks \citep{MIST2016I,MIST2016II,MESA2011,MESA2013,MESA2015,MESA2018}) according to its mass. All the MIST tracks in Fig. \ref{fig:delta} were calculated with zero rotational velocity and an initial metallicity of 0.02. The above results indicate that the current stellar properties of the BSS in WOCS 1007 are similar to that of single $\delta$ Scuti stars.

We also compared the properties of the primary star in WOCS 1007 with binary $\delta$ Scuti stars. The correlation between the dominant pulsation and the orbital periods of $\delta$ Scuti stars in eclipsing binaries are well established \citep{2017MNRAS.465.1181L,2020AcA....70..265L}. For the comparison, we used detached eclipsing systems and oscillating eclipsing Algol-type binary (oEA) systems. 
We considered $\delta$ Scuti stars in detached systems from \cite{2017MNRAS.465.1181L} and \cite{2020AcA....70..265L}, and semi-detached or oEA systems (taken from \citet{2017MNRAS.465.1181L}). In Fig. \ref{fig:relations}, the BSS was placed on the P$_{\rm orb}$--P$_{\rm pul}$ (top panel) and P$_{\rm pul}$--log g (bottom panel) diagrams. In the top panel, the BSS is located below the fitted lines but within the region occupied by the detached and oEA binaries. In the bottom panel, it is consistent with both the distributions as well as the empirical relations. Therefore the properties of the WOCS 1007 system are consistent with the properties of binary systems hosting $\delta$ Scuti stars.

The system WOCS 4003 is reported as an SB1 \citep{Geller2015} but shows $\beta$ Lyrae-type light curve. We observed certain features in the light curve, such as the O'Connell effect and small dips equally spaced throughout the light curve. The orbital period estimate from this study is in good agreement with \citet{2009A&A...503..165Y}. As there was no proper constraint on the system's configuration, we modelled the system using three different models (W UMa type overcontact binary, non-thermal overcontact binary, and semi-detached binary with primary filling Roche lobe). Our PHOEBE fitting results from all three models were different from each other. Even though the radius of the primary and effective temperature of the secondary agreed with the SED values, the secondary radius from all three models was significantly higher than the SED solution, leading to an over-luminous system. Overall, we were unable to obtain satisfactory parameters for the components.
In all three models, the inclination of the system was determined to be quite low, suggesting the system to be exhibiting a grazing eclipse. Fourier analysis showed that the light curve is affected by both eclipses and ellipsoidal variability. Along with the large eclipses with a period of 0.44 d, the light curve also shows periodic dips of 1.1 d. Though we excluded contamination from nearby sources by checking their light curves, we are unable to confirm the presence of a tertiary component in the system. 

WOCS 4006 was classified as an SB1 system with $\delta$ Scuti pulsations. We did not detect any eclipses though $\delta$ Scuti pulsations were detected. Similar to WOCS 1007, this BSS also appears to have been rejuvenated after the mass transfer. Upon comparing the SED parameters with the single star evolutionary tracks, the star's mass was around 1.6 M$_{\sun}$. We detect 29 frequencies in the amplitude spectrum, higher than the number of frequencies detected in WOCS1007, though not as high as the 41, which was reported in \citet{Bruntt2007}.

We did not detect eclipses in the light curves for two more BSSs that are found to be SB1 systems: WOCS 5005 and WOCS 1025.
This could be due to the low inclination of the systems. In the literature, WOCS 5005 (4913 d) and WOCS 1025 (1154 d) were reported as long-period binaries with an LM WD and a WD as the secondary companions, respectively. As LM WD cannot form through single star evolution within the Hubble time, \citet{2021MNRAS.507.2373P} speculated that systems like WOCS 5005 could either be formed due to dynamical interactions \citep{Khurana2022arXiv220906231K} or have the LM WD in the close inner binary of a tertiary system. As we did not detect any eclipses in the light curves of these systems, we can confirm the absence of short-period inner binary with a high inclination that gives rise to eclipses or ellipsoidal variations. We note that this study does not rule out the presence of a binary component with slightly larger periods and/or low inclinations. Therefore, this study shows that TESS/K2 light curves are helpful in detecting close companions of BSSs in highly-inclined orbits. If detected, they provide essential constraints on the properties as well as the formation of the binary.

\section{Summary}

This study presents a light curve analysis of select BSSs in M67 using the TESS and K2 data. We summarise the results below:

\begin{itemize}
   \item All the sources studied are known to be SB1 systems suggesting a low luminosity companion to the BSS primary.
    \item We analysed the light curve for the BSS WOCS 1007 for the first time. Our analysis using PHOEBE confirms 
    WOCS 1007 to have an LM WD companion. We estimate P = 4.212$\pm$0.041 d and $e$ = 0.206$\pm$0.002  The masses of companions are estimated to be 1.95$\pm$0.26 and 0.22$\pm$0.05 M$_{\sun}$ with radii of 2.54$\pm$0.81 and 0.078$\pm$0.027 R$_{\sun}$, respectively. The temperature of the LM WD is found to be 14300$\pm$1100 K. The estimated mass of the BSS is similar to the mass estimated from isochrones.
    \item We detect the $\delta$ Scuti pulsations in the BSS of WOCS 1007 (19 frequencies). The observed light curve is primarily due to ellipsoidal variations in the system, suggestive of a slight deformation in the BSS, not enough to inhibit the $\delta$ Scuti pulsations.
    \item The WOCS 1007 is likely to be formed by efficient mass transfer in a close binary, leaving behind an LM WD. The increased mass has pushed it to within the instability strip. This BSS is found to be similar to other $\delta$ Scuti stars in eclipsing binaries. 
    \item WOCS 4003 is modelled with various options, but PHOEBE results suggest an over-luminous system. The best model shows a grazing eclipse, but ellipsoidal variation is dominant in the light curve. 
    \item Apart from the 0.44 d period eclipses in WOCS 4003, we found smaller eclipses with a period of 1.1 d. There may be a close tertiary, making it a compact triple system. WOCS 4003 is a complex system, and multi-wavelength light curves are needed to identify and characterise the individual components.
    \item WOCS 4006 is a known $\delta$ Scuti star, and we detect 29 pulsation frequencies, whereas no eclipses are detected. No eclipses or pulsations are detected in the case of WOCS 5005 and WOCS 1025. We confirm the absence of any short-period inner binary with high inclination in these BSSs. This method cannot rule out a slightly distant inner binary companion or a close companion in an almost face-on inclination.  
\end{itemize}

In conclusion, our study provides additional evidence supporting the mass transfer mechanism of BSS formation in lower-density regions. It also demonstrates that the light curve analysis of BSSs in open clusters, along with the multi-wavelength SED analysis, can address the nature and formation pathways of the BSSs. We plan to use this technique to analyse BSSs in other clusters. From the identification and analysis of WOCS 1007, we have shown that light curve analysis is a valuable technique for studying such short-period post-mass transfer objects. Combining this technique and the one used by \cite{2018MNRAS.474.4322M} will help identify both short and long-period post-mass transfer systems. The short-period detached nature of WOCS 1007 BSS, along with its $\delta$ Scuti pulsations, makes it a prime candidate for a further detailed analysis using asteroseismology, as this is the first such a system identified in an open cluster. Also, only 44 such systems are known in the field \citep{2020AcA....70..265L}. Considering WOCS 1007 has already interacted, such a study will not only provide information on the effects of binarity on pulsations but also help understand the influence of mass transfer on pulsations.

\section*{Acknowledgements}
We thank the referee for constructive comments and suggestions.
AS acknowledges support from a SERB Power Fellowship. 
DMB gratefully acknowledges a senior postdoctoral fellowship from the Research Foundation Flanders (FWO) with the grant agreement no. 1286521N.
VJ thanks the Alexander von Humboldt Foundation for their support.
This paper includes data collected by the \textit{TESS} mission, which are publicly available from the Mikulski Archive for Space Telescopes (MAST). Funding for the \textit{TESS} mission is provided by NASA's Science Mission Directorate. 
This paper includes data collected by the \textit{Kepler} mission and obtained from the MAST data archive at the Space Telescope Science Institute (STScI). Funding for the \textit{Kepler} mission is provided by the NASA Science Mission Directorate. STScI is operated by the Association of Universities for Research in Astronomy, Inc., under NASA contract NAS 5–26555. 
This work presents results from the European Space Agency (ESA) space mission \textit{Gaia}. \textit{Gaia} data are being processed by the \textit{Gaia} Data Processing and Analysis Consortium (DPAC) (https://www.cosmos.esa.int/gaia).
This work and the Pythia code used in this work make use of Python packages (Matplotlib \citep{Hunter:2007}, Numpy \citep{harris2020array} and PyMC3 \citep{2015arXiv150708050S}). This research made use of \textsf{exoplanet} \citep{exoplanet:exoplanet} and its dependencies \citep{exoplanet:agol20,exoplanet:astropy13, exoplanet:astropy18, exoplanet:kipping13, exoplanet:luger18,
exoplanet:theano}. 
This research has made use of the SIMBAD database, operated at CDS, Strasbourg, France \citep{2000A&AS..143....9W}. 

\section*{Data Availability}

All the \textit{TESS} and \textit{K2} data can be obtained through the Mikulski Archive
for Space Telescopes (MAST) database. 
\bibliographystyle{mnras}
\bibliography{final_solution} 

\begin{thebibliography}{}
\makeatletter
\relax
\def\mn@urlcharsother{\let\do\@makeother \do\$\do\&\do\#\do\^\do\_\do\%\do\~}
\def\mn@doi{\begingroup\mn@urlcharsother \@ifnextchar [ {\mn@doi@}
  {\mn@doi@[]}}
\def\mn@doi@[#1]#2{\def\@tempa{#1}\ifx\@tempa\@empty \href
  {http://dx.doi.org/#2} {doi:#2}\else \href {http://dx.doi.org/#2} {#1}\fi
  \endgroup}
\def\mn@eprint#1#2{\mn@eprint@#1:#2::\@nil}
\def\mn@eprint@arXiv#1{\href {http://arxiv.org/abs/#1} {{\tt arXiv:#1}}}
\def\mn@eprint@dblp#1{\href {http://dblp.uni-trier.de/rec/bibtex/#1.xml}
  {dblp:#1}}
\def\mn@eprint@#1:#2:#3:#4\@nil{\def\@tempa {#1}\def\@tempb {#2}\def\@tempc
  {#3}\ifx \@tempc \@empty \let \@tempc \@tempb \let \@tempb \@tempa \fi \ifx
  \@tempb \@empty \def\@tempb {arXiv}\fi \@ifundefined
  {mn@eprint@\@tempb}{\@tempb:\@tempc}{\expandafter \expandafter \csname
  mn@eprint@\@tempb\endcsname \expandafter{\@tempc}}}

\bibitem[\protect\citeauthoryear{{Agol}, {Luger}  \& {Foreman-Mackey}}{{Agol}
  et~al.}{2020}]{exoplanet:agol20}
{Agol} E.,  {Luger} R.,   {Foreman-Mackey} D.,  2020, \mn@doi [\aj]
  {10.3847/1538-3881/ab4fee}, \href
  {https://ui.adsabs.harvard.edu/abs/2020AJ....159..123A} {159, 123}

\bibitem[\protect\citeauthoryear{{Aigrain}, {Parviainen}  \& {Pope}}{{Aigrain}
  et~al.}{2016}]{2016MNRAS.459.2408A}
{Aigrain} S.,  {Parviainen} H.,   {Pope} B.~J.~S.,  2016, \mn@doi [\mnras]
  {10.1093/mnras/stw706}, \href
  {https://ui.adsabs.harvard.edu/abs/2016MNRAS.459.2408A} {459, 2408}

\bibitem[\protect\citeauthoryear{{Allen} \& {Strom}}{{Allen} \&
  {Strom}}{1995}]{1995AJ....109.1379A}
{Allen} L.~E.,  {Strom} K.~M.,  1995, \mn@doi [\aj] {10.1086/117370}, \href
  {https://ui.adsabs.harvard.edu/abs/1995AJ....109.1379A} {109, 1379}

\bibitem[\protect\citeauthoryear{{Astropy Collaboration} et~al.,}{{Astropy
  Collaboration} et~al.}{2013}]{exoplanet:astropy13}
{Astropy Collaboration} et~al., 2013, \mn@doi [\aap]
  {10.1051/0004-6361/201322068}, \href
  {http://adsabs.harvard.edu/abs/2013A%26A...558A..33A} {558, A33}

\bibitem[\protect\citeauthoryear{{Astropy Collaboration} et~al.,}{{Astropy
  Collaboration} et~al.}{2018}]{exoplanet:astropy18}
{Astropy Collaboration} et~al., 2018, \mn@doi [\aj] {10.3847/1538-3881/aabc4f},
  \href {http://adsabs.harvard.edu/abs/2018AJ....156..123A} {156, 123}

\bibitem[\protect\citeauthoryear{{Beech}}{{Beech}}{1985}]{1985Ap&SS.117...69B}
{Beech} M.,  1985, \mn@doi [\apss] {10.1007/BF00660911}, \href
  {https://ui.adsabs.harvard.edu/abs/1985Ap&SS.117...69B} {117, 69}

\bibitem[\protect\citeauthoryear{{Bertelli Motta}, {Pasquali}, {Caffau}  \&
  {Grebel}}{{Bertelli Motta} et~al.}{2018}]{2018MNRAS.480.4314B}
{Bertelli Motta} C.,  {Pasquali} A.,  {Caffau} E.,   {Grebel} E.~K.,  2018,
  \mn@doi [\mnras] {10.1093/mnras/sty2147}, \href
  {https://ui.adsabs.harvard.edu/abs/2018MNRAS.480.4314B} {480, 4314}

\bibitem[\protect\citeauthoryear{{Bowman} \& {Kurtz}}{{Bowman} \&
  {Kurtz}}{2018}]{bowman2018}
{Bowman} D.~M.,  {Kurtz} D.~W.,  2018, \mn@doi [\mnras] {10.1093/mnras/sty449},
  \href {https://ui.adsabs.harvard.edu/abs/2018MNRAS.476.3169B} {476, 3169}

\bibitem[\protect\citeauthoryear{{Bowman} \& {Michielsen}}{{Bowman} \&
  {Michielsen}}{2021}]{2021A&A...656A.158B}
{Bowman} D.~M.,  {Michielsen} M.,  2021, \mn@doi [\aap]
  {10.1051/0004-6361/202141726}, \href
  {https://ui.adsabs.harvard.edu/abs/2021A&A...656A.158B} {656, A158}

\bibitem[\protect\citeauthoryear{{Bowman}, {Johnston}, {Tkachenko},
  {Mkrtichian}, {Gunsriwiwat}  \& {Aerts}}{{Bowman}
  et~al.}{2019}]{2019ApJ...883L..26B}
{Bowman} D.~M.,  {Johnston} C.,  {Tkachenko} A.,  {Mkrtichian} D.~E.,
  {Gunsriwiwat} K.,   {Aerts} C.,  2019, \mn@doi [\apjl]
  {10.3847/2041-8213/ab3fb2}, \href
  {https://ui.adsabs.harvard.edu/abs/2019ApJ...883L..26B} {883, L26}

\bibitem[\protect\citeauthoryear{{Breger}}{{Breger}}{2000}]{breger2000}
{Breger} M.,  2000, in {Breger} M.,  {Montgomery} M.,  eds,  Astronomical
  Society of the Pacific Conference Series Vol. 210, Delta Scuti and Related
  Stars. p.~3

\bibitem[\protect\citeauthoryear{{Breger} et~al.,}{{Breger}
  et~al.}{1993}]{1993A&A...271..482B}
{Breger} M.,  et~al., 1993, \aap, \href
  {https://ui.adsabs.harvard.edu/abs/1993A&A...271..482B} {271, 482}

\bibitem[\protect\citeauthoryear{{Bruntt} et~al.,}{{Bruntt}
  et~al.}{2007}]{Bruntt2007}
{Bruntt} H.,  et~al., 2007, \mn@doi [\mnras]
  {10.1111/j.1365-2966.2007.11865.x}, \href
  {http://adsabs.harvard.edu/abs/2007MNRAS.378.1371B} {378, 1371}

\bibitem[\protect\citeauthoryear{{Choi}, {Dotter}, {Conroy}, {Cantiello},
  {Paxton}  \& {Johnson}}{{Choi} et~al.}{2016}]{MIST2016I}
{Choi} J.,  {Dotter} A.,  {Conroy} C.,  {Cantiello} M.,  {Paxton} B.,
  {Johnson} B.~D.,  2016, \mn@doi [\apj] {10.3847/0004-637X/823/2/102}, \href
  {https://ui.adsabs.harvard.edu/abs/2016ApJ...823..102C} {823, 102}

\bibitem[\protect\citeauthoryear{{Dal} \& {Sipahi}}{{Dal} \&
  {Sipahi}}{2013}]{2013PASA...30...16D}
{Dal} H.~A.,  {Sipahi} E.,  2013, \mn@doi [\pasa] {10.1017/pasa.2012.016},
  \href {https://ui.adsabs.harvard.edu/abs/2013PASA...30...16D} {30, e016}

\bibitem[\protect\citeauthoryear{{Deng}, {Chen}, {Liu}  \& {Chen}}{{Deng}
  et~al.}{1999}]{Deng99}
{Deng} L.,  {Chen} R.,  {Liu} X.~S.,   {Chen} J.~S.,  1999, \mn@doi [\apj]
  {10.1086/307832}, \href {http://adsabs.harvard.edu/abs/1999ApJ...524..824D}
  {524, 824}

\bibitem[\protect\citeauthoryear{{Dotter}}{{Dotter}}{2016}]{MIST2016II}
{Dotter} A.,  2016, \mn@doi [\apjs] {10.3847/0067-0049/222/1/8}, \href
  {https://ui.adsabs.harvard.edu/abs/2016ApJS..222....8D} {222, 8}

\bibitem[\protect\citeauthoryear{{Ebbighausen}}{{Ebbighausen}}{1940}]{1940ApJ....91..244E}
{Ebbighausen} E.~G.,  1940, \mn@doi [\apj] {10.1086/144160}, \href
  {https://ui.adsabs.harvard.edu/abs/1940ApJ....91..244E} {91, 244}

\bibitem[\protect\citeauthoryear{{Ferreira Lopes}, {D{\'e}k{\'a}ny}, {Catelan},
  {Cross}, {Angeloni}, {Le{\~a}o}  \& {De Medeiros}}{{Ferreira Lopes}
  et~al.}{2015}]{2015A&A...573A.100F}
{Ferreira Lopes} C.~E.,  {D{\'e}k{\'a}ny} I.,  {Catelan} M.,  {Cross} N.~J.~G.,
   {Angeloni} R.,  {Le{\~a}o} I.~C.,   {De Medeiros} J.~R.,  2015, \mn@doi
  [\aap] {10.1051/0004-6361/201423793}, \href
  {https://ui.adsabs.harvard.edu/abs/2015A&A...573A.100F} {573, A100}

\bibitem[\protect\citeauthoryear{{Foreman-Mackey}, {Hogg}, {Lang}  \&
  {Goodman}}{{Foreman-Mackey} et~al.}{2013}]{emcee}
{Foreman-Mackey} D.,  {Hogg} D.~W.,  {Lang} D.,   {Goodman} J.,  2013, \mn@doi
  [\pasp] {10.1086/670067}, \href
  {https://ui.adsabs.harvard.edu/abs/2013PASP..125..306F} {125, 306}

\bibitem[\protect\citeauthoryear{Foreman-Mackey, Luger, Czekala, Agol,
  Price-Whelan, Brandt, Barclay  \& Bouma}{Foreman-Mackey
  et~al.}{2020}]{exoplanet:exoplanet}
Foreman-Mackey D.,  Luger R.,  Czekala I.,  Agol E.,  Price-Whelan A.,  Brandt
  T.~D.,  Barclay T.,   Bouma L.,  2020, exoplanet-dev/exoplanet v0.4.0,
  \mn@doi{10.5281/zenodo.1998447}, \url
  {https://doi.org/10.5281/zenodo.1998447}

\bibitem[\protect\citeauthoryear{{Gaia Collaboration}, {Vallenari, A.}, {Brown,
  A.G.A.}, {Prusti, T.}  \& {et al.}}{{Gaia Collaboration}
  et~al.}{2022}]{GaiaDR3_2022}
{Gaia Collaboration} {Vallenari, A.} {Brown, A.G.A.} {Prusti, T.}  {et al.}
  2022, \mn@doi [A\&A] {10.1051/0004-6361/202243940}

\bibitem[\protect\citeauthoryear{{Geller}, {Latham}  \& {Mathieu}}{{Geller}
  et~al.}{2015}]{Geller2015}
{Geller} A.~M.,  {Latham} D.~W.,   {Mathieu} R.~D.,  2015, \mn@doi [\aj]
  {10.1088/0004-6256/150/3/97}, \href
  {https://ui.adsabs.harvard.edu/abs/2015AJ....150...97G} {150, 97}

\bibitem[\protect\citeauthoryear{{Gilliland} \& {Brown}}{{Gilliland} \&
  {Brown}}{1992}]{Gilliland1992}
{Gilliland} R.~L.,  {Brown} T.~M.,  1992, \mn@doi [\aj] {10.1086/116203}, \href
  {http://adsabs.harvard.edu/abs/1992AJ....103.1945G} {103, 1945}

\bibitem[\protect\citeauthoryear{{Gilliland} et~al.,}{{Gilliland}
  et~al.}{1991}]{1991AJ....101..541G}
{Gilliland} R.~L.,  et~al., 1991, \mn@doi [\aj] {10.1086/115703}, \href
  {https://ui.adsabs.harvard.edu/abs/1991AJ....101..541G} {101, 541}

\bibitem[\protect\citeauthoryear{{Goodman} \& {Weare}}{{Goodman} \&
  {Weare}}{2010}]{goodman}
{Goodman} J.,  {Weare} J.,  2010, \mn@doi [Communications in Applied
  Mathematics and Computational Science] {10.2140/camcos.2010.5.65}, \href
  {https://ui.adsabs.harvard.edu/abs/2010CAMCS...5...65G} {5, 65}

\bibitem[\protect\citeauthoryear{{Gosnell}, {Mathieu}, {Geller}, {Sills},
  {Leigh}  \& {Knigge}}{{Gosnell} et~al.}{2015}]{Gosnell2015}
{Gosnell} N.~M.,  {Mathieu} R.~D.,  {Geller} A.~M.,  {Sills} A.,  {Leigh} N.,
  {Knigge} C.,  2015, \mn@doi [\apj] {10.1088/0004-637X/814/2/163}, \href
  {https://ui.adsabs.harvard.edu/abs/2015ApJ...814..163G} {814, 163}

\bibitem[\protect\citeauthoryear{{Gosnell}, {Leiner}, {Mathieu}, {Geller},
  {Knigge}, {Sills}  \& {Leigh}}{{Gosnell} et~al.}{2019}]{Gosnell2019}
{Gosnell} N.~M.,  {Leiner} E.~M.,  {Mathieu} R.~D.,  {Geller} A.~M.,  {Knigge}
  C.,  {Sills} A.,   {Leigh} N. W.~C.,  2019, \mn@doi [\apj]
  {10.3847/1538-4357/ab4273}, \href
  {https://ui.adsabs.harvard.edu/abs/2019ApJ...885...45G} {885, 45}

\bibitem[\protect\citeauthoryear{{Handler} et~al.,}{{Handler}
  et~al.}{2020}]{2020NatAs...4..684H}
{Handler} G.,  et~al., 2020, \mn@doi [Nature Astronomy]
  {10.1038/s41550-020-1035-1}, \href
  {https://ui.adsabs.harvard.edu/abs/2020NatAs...4..684H} {4, 684}

\bibitem[\protect\citeauthoryear{Harris et~al.,}{Harris
  et~al.}{2020}]{harris2020array}
Harris C.~R.,  et~al., 2020, \mn@doi [Nature] {10.1038/s41586-020-2649-2}, 585,
  357

\bibitem[\protect\citeauthoryear{{Henry} \& {Kaye}}{{Henry} \&
  {Kaye}}{1999}]{1999IBVS.4684....1H}
{Henry} G.~W.,  {Kaye} A.~B.,  1999, Information Bulletin on Variable Stars,
  \href {https://ui.adsabs.harvard.edu/abs/1999IBVS.4684....1H} {4684, 1}

\bibitem[\protect\citeauthoryear{{Howell} et~al.,}{{Howell}
  et~al.}{2014}]{2014PASP..126..398H}
{Howell} S.~B.,  et~al., 2014, \mn@doi [\pasp] {10.1086/676406}, \href
  {https://ui.adsabs.harvard.edu/abs/2014PASP..126..398H} {126, 398}

\bibitem[\protect\citeauthoryear{Hunter}{Hunter}{2007}]{Hunter:2007}
Hunter J.~D.,  2007, \mn@doi [Computing in Science \& Engineering]
  {10.1109/MCSE.2007.55}, 9, 90

\bibitem[\protect\citeauthoryear{{Jadhav} \& {Subramaniam}}{{Jadhav} \&
  {Subramaniam}}{2021}]{Jadhav2021MNRAS.507.1699J}
{Jadhav} V.~V.,  {Subramaniam} A.,  2021, \mn@doi [\mnras]
  {10.1093/mnras/stab2264}, \href
  {https://ui.adsabs.harvard.edu/abs/2021MNRAS.507.1699J} {507, 1699}

\bibitem[\protect\citeauthoryear{{Jadhav}, {Sindhu}  \& {Subramaniam}}{{Jadhav}
  et~al.}{2019}]{Jadhav2019}
{Jadhav} V.~V.,  {Sindhu} N.,   {Subramaniam} A.,  2019, \mn@doi [\apj]
  {10.3847/1538-4357/ab4b43}, \href
  {https://ui.adsabs.harvard.edu/abs/2019ApJ...886...13J} {886, 13}

\bibitem[\protect\citeauthoryear{{Jenkins} et~al.,}{{Jenkins}
  et~al.}{2010}]{2010ApJ...713L..87J}
{Jenkins} J.~M.,  et~al., 2010, \mn@doi [\apjl] {10.1088/2041-8205/713/2/L87},
  \href {https://ui.adsabs.harvard.edu/abs/2010ApJ...713L..87J} {713, L87}

\bibitem[\protect\citeauthoryear{{Jenkins} et~al.,}{{Jenkins}
  et~al.}{2016}]{2016SPIE.9913E..3EJ}
{Jenkins} J.~M.,  et~al., 2016, in {Chiozzi} G.,  {Guzman} J.~C.,  eds,
  Society of Photo-Optical Instrumentation Engineers (SPIE) Conference Series
  Vol. 9913, Software and Cyberinfrastructure for Astronomy IV. p. 99133E,
  \mn@doi{10.1117/12.2233418}

\bibitem[\protect\citeauthoryear{{Khurana}, {Chawla}  \&
  {Chatterjee}}{{Khurana} et~al.}{2022}]{Khurana2022arXiv220906231K}
{Khurana} A.,  {Chawla} C.,   {Chatterjee} S.,  2022, arXiv e-prints, \href
  {https://ui.adsabs.harvard.edu/abs/2022arXiv220906231K} {p. arXiv:2209.06231}

\bibitem[\protect\citeauthoryear{{Kippenhahn} \& {Weigert}}{{Kippenhahn} \&
  {Weigert}}{1967}]{Kippenhahn1967}
{Kippenhahn} R.,  {Weigert} A.,  1967, \zap, \href
  {https://ui.adsabs.harvard.edu/abs/1967ZA.....65..251K} {65, 251}

\bibitem[\protect\citeauthoryear{{Kipping}}{{Kipping}}{2013}]{exoplanet:kipping13}
{Kipping} D.~M.,  2013, \mn@doi [\mnras] {10.1093/mnras/stt1435}, \href
  {http://adsabs.harvard.edu/abs/2013MNRAS.435.2152K} {435, 2152}

\bibitem[\protect\citeauthoryear{{Knigge}, {Dieball}, {Ma{\'{\i}}z
  Apell{\'a}niz}, {Long}, {Zurek}  \& {Shara}}{{Knigge}
  et~al.}{2008}]{Knigge2008}
{Knigge} C.,  {Dieball} A.,  {Ma{\'{\i}}z Apell{\'a}niz} J.,  {Long} K.~S.,
  {Zurek} D.~R.,   {Shara} M.~M.,  2008, \mn@doi [\apj] {10.1086/589987}, \href
  {http://adsabs.harvard.edu/abs/2008ApJ...683.1006K} {683, 1006}

\bibitem[\protect\citeauthoryear{{Lacy}}{{Lacy}}{1992}]{1992ASPC...32..152L}
{Lacy} C.~H.,  1992, in {McAlister} H.~A.,  {Hartkopf} W.~I.,  eds,
  Astronomical Society of the Pacific Conference Series Vol. 32, IAU Colloq.
  135: Complementary Approaches to Double and Multiple Star Research. p.~152

\bibitem[\protect\citeauthoryear{{Latham} \& {Milone}}{{Latham} \&
  {Milone}}{1996}]{Latham1996}
{Latham} D.~W.,  {Milone} A.~A.~E.,  1996, in {Milone} E.~F.,  {Mermilliod}
  J.-C.,  eds,  Astronomical Society of the Pacific Conference Series Vol. 90,
  The origins, evolution, and destinies of binary stars in clusters. pp
  385--387

\bibitem[\protect\citeauthoryear{{Lauterborn}}{{Lauterborn}}{1970}]{Lauterborn1970}
{Lauterborn} D.,  1970, \aap, \href
  {https://ui.adsabs.harvard.edu/abs/1970A&A.....7..150L} {7, 150}

\bibitem[\protect\citeauthoryear{{Leiner}, {Mathieu}, {Vanderburg}, {Gosnell}
  \& {Smith}}{{Leiner} et~al.}{2019}]{Leiner2019}
{Leiner} E.,  {Mathieu} R.~D.,  {Vanderburg} A.,  {Gosnell} N.~M.,   {Smith}
  J.~C.,  2019, \mn@doi [\apj] {10.3847/1538-4357/ab2bf8}, \href
  {https://ui.adsabs.harvard.edu/abs/2019ApJ...881...47L} {881, 47}

\bibitem[\protect\citeauthoryear{{Lenz} \& {Breger}}{{Lenz} \&
  {Breger}}{2014}]{2014ascl.soft07009L}
{Lenz} P.,  {Breger} M.,  2014, {Period04: Statistical analysis of large
  astronomical time series}, Astrophysics Source Code Library, record
  ascl:1407.009 (\mn@eprint {ascl} {1407.009})

\bibitem[\protect\citeauthoryear{{Liakos}}{{Liakos}}{2020}]{2020AcA....70..265L}
{Liakos} A.,  2020, \mn@doi [\actaa] {10.32023/0001-5237/70.4.2}, \href
  {https://ui.adsabs.harvard.edu/abs/2020AcA....70..265L} {70, 265}

\bibitem[\protect\citeauthoryear{{Liakos} \& {Niarchos}}{{Liakos} \&
  {Niarchos}}{2017}]{2017MNRAS.465.1181L}
{Liakos} A.,  {Niarchos} P.,  2017, \mn@doi [\mnras] {10.1093/mnras/stw2756},
  \href {https://ui.adsabs.harvard.edu/abs/2017MNRAS.465.1181L} {465, 1181}

\bibitem[\protect\citeauthoryear{{Lu}, {Deng}  \& {Zhang}}{{Lu}
  et~al.}{2010}]{Lu2010}
{Lu} P.,  {Deng} L.~C.,   {Zhang} X.~B.,  2010, \mn@doi [\mnras]
  {10.1111/j.1365-2966.2010.17356.x}, \href
  {https://ui.adsabs.harvard.edu/abs/2010MNRAS.409.1013L} {409, 1013}

\bibitem[\protect\citeauthoryear{{Luger}, {Agol}, {Foreman-Mackey}, {Fleming},
  {Lustig-Yaeger}  \& {Deitrick}}{{Luger} et~al.}{2019}]{exoplanet:luger18}
{Luger} R.,  {Agol} E.,  {Foreman-Mackey} D.,  {Fleming} D.~P.,
  {Lustig-Yaeger} J.,   {Deitrick} R.,  2019, \mn@doi [\aj]
  {10.3847/1538-3881/aae8e5}, \href
  {http://adsabs.harvard.edu/abs/2019AJ....157...64L} {157, 64}

\bibitem[\protect\citeauthoryear{{Manteiga}, {Pickles}  \&
  {Martinez-Roger}}{{Manteiga} et~al.}{1989}]{Manteiga1989}
{Manteiga} M.,  {Pickles} A.~J.,   {Martinez-Roger} C.,  1989, \aap, \href
  {https://ui.adsabs.harvard.edu/abs/1989A&A...210...66M} {210, 66}

\bibitem[\protect\citeauthoryear{{Mathys}}{{Mathys}}{1991}]{1991A&A...245..467M}
{Mathys} G.,  1991, \aap, \href
  {https://ui.adsabs.harvard.edu/abs/1991A&A...245..467M} {245, 467}

\bibitem[\protect\citeauthoryear{{McCrea}}{{McCrea}}{1964}]{McCrea1964}
{McCrea} W.~H.,  1964, \mn@doi [\mnras] {10.1093/mnras/128.2.147}, \href
  {https://ui.adsabs.harvard.edu/#abs/1964MNRAS.128..147M} {128, 147}

\bibitem[\protect\citeauthoryear{{Milone} \& {Latham}}{{Milone} \&
  {Latham}}{1992}]{1992IAUS..151..475M}
{Milone} A.~A.~E.,  {Latham} D.~W.,  1992, {The Blue Straggler F:190 - a Case
  for Mass Transfer}

\bibitem[\protect\citeauthoryear{{Milone}, {Latham}, {Kurucz}  \&
  {Morse}}{{Milone} et~al.}{1991}]{1991ASPC...13..424M}
{Milone} A. A.~E.,  {Latham} D.~W.,  {Kurucz} R.~L.,   {Morse} J.~A.,  1991, in
  {Janes} K.,  ed.,  Astronomical Society of the Pacific Conference Series Vol.
  13, The Formation and Evolution of Star Clusters. pp 424--426

\bibitem[\protect\citeauthoryear{{Morris}}{{Morris}}{1985}]{1985ApJ...295..143M}
{Morris} S.~L.,  1985, \mn@doi [\apj] {10.1086/163359}, \href
  {https://ui.adsabs.harvard.edu/abs/1985ApJ...295..143M} {295, 143}

\bibitem[\protect\citeauthoryear{{Morris} \& {Naftilan}}{{Morris} \&
  {Naftilan}}{1993}]{1993ApJ...419..344M}
{Morris} S.~L.,  {Naftilan} S.~A.,  1993, \mn@doi [\apj] {10.1086/173488},
  \href {https://ui.adsabs.harvard.edu/abs/1993ApJ...419..344M} {419, 344}

\bibitem[\protect\citeauthoryear{{Murphy}, {Moe}, {Kurtz}, {Bedding},
  {Shibahashi}  \& {Boffin}}{{Murphy} et~al.}{2018}]{2018MNRAS.474.4322M}
{Murphy} S.~J.,  {Moe} M.,  {Kurtz} D.~W.,  {Bedding} T.~R.,  {Shibahashi} H.,
   {Boffin} H. M.~J.,  2018, \mn@doi [\mnras] {10.1093/mnras/stx3049}, \href
  {https://ui.adsabs.harvard.edu/abs/2018MNRAS.474.4322M} {474, 4322}

\bibitem[\protect\citeauthoryear{{Murphy}, {Hey}, {Van Reeth}  \&
  {Bedding}}{{Murphy} et~al.}{2019}]{murphy2019}
{Murphy} S.~J.,  {Hey} D.,  {Van Reeth} T.,   {Bedding} T.~R.,  2019, \mn@doi
  [\mnras] {10.1093/mnras/stz590}, \href
  {https://ui.adsabs.harvard.edu/abs/2019MNRAS.485.2380M} {485, 2380}

\bibitem[\protect\citeauthoryear{{Naoz} \& {Fabrycky}}{{Naoz} \&
  {Fabrycky}}{2014}]{Naoz2014}
{Naoz} S.,  {Fabrycky} D.~C.,  2014, \mn@doi [\apj]
  {10.1088/0004-637X/793/2/137}, \href
  {https://ui.adsabs.harvard.edu/#abs/2014ApJ...793..137N} {793, 137}

\bibitem[\protect\citeauthoryear{{Nascimbeni} et~al.,}{{Nascimbeni}
  et~al.}{2016}]{2016MNRAS.463.4210N}
{Nascimbeni} V.,  et~al., 2016, \mn@doi [\mnras] {10.1093/mnras/stw2313}, \href
  {https://ui.adsabs.harvard.edu/abs/2016MNRAS.463.4210N} {463, 4210}

\bibitem[\protect\citeauthoryear{{Paczy{\'n}ski}, {Szczygie{\l}}, {Pilecki}  \&
  {Pojma{\'n}ski}}{{Paczy{\'n}ski} et~al.}{2006}]{2006MNRAS.368.1311P}
{Paczy{\'n}ski} B.,  {Szczygie{\l}} D.~M.,  {Pilecki} B.,   {Pojma{\'n}ski} G.,
   2006, \mn@doi [\mnras] {10.1111/j.1365-2966.2006.10223.x}, \href
  {https://ui.adsabs.harvard.edu/abs/2006MNRAS.368.1311P} {368, 1311}

\bibitem[\protect\citeauthoryear{{Pandey}, {Subramaniam}  \& {Jadhav}}{{Pandey}
  et~al.}{2021}]{2021MNRAS.507.2373P}
{Pandey} S.,  {Subramaniam} A.,   {Jadhav} V.~V.,  2021, \mn@doi [\mnras]
  {10.1093/mnras/stab2308}, \href
  {https://ui.adsabs.harvard.edu/abs/2021MNRAS.507.2373P} {507, 2373}

\bibitem[\protect\citeauthoryear{{Paxton}, {Bildsten}, {Dotter}, {Herwig},
  {Lesaffre}  \& {Timmes}}{{Paxton} et~al.}{2011}]{MESA2011}
{Paxton} B.,  {Bildsten} L.,  {Dotter} A.,  {Herwig} F.,  {Lesaffre} P.,
  {Timmes} F.,  2011, \mn@doi [\apjs] {10.1088/0067-0049/192/1/3}, \href
  {https://ui.adsabs.harvard.edu/abs/2011ApJS..192....3P} {192, 3}

\bibitem[\protect\citeauthoryear{{Paxton} et~al.,}{{Paxton}
  et~al.}{2013}]{MESA2013}
{Paxton} B.,  et~al., 2013, \mn@doi [\apjs] {10.1088/0067-0049/208/1/4}, \href
  {https://ui.adsabs.harvard.edu/abs/2013ApJS..208....4P} {208, 4}

\bibitem[\protect\citeauthoryear{{Paxton} et~al.,}{{Paxton}
  et~al.}{2015}]{MESA2015}
{Paxton} B.,  et~al., 2015, \mn@doi [\apjs] {10.1088/0067-0049/220/1/15}, \href
  {https://ui.adsabs.harvard.edu/abs/2015ApJS..220...15P} {220, 15}

\bibitem[\protect\citeauthoryear{{Paxton} et~al.,}{{Paxton}
  et~al.}{2018}]{MESA2018}
{Paxton} B.,  et~al., 2018, \mn@doi [\apjs] {10.3847/1538-4365/aaa5a8}, \href
  {https://ui.adsabs.harvard.edu/abs/2018ApJS..234...34P} {234, 34}

\bibitem[\protect\citeauthoryear{{Perets}}{{Perets}}{2015}]{Perets2015}
{Perets} H.~B.,  2015, in Ecology of Blue Straggler Stars. p.~251

\bibitem[\protect\citeauthoryear{{Perets} \& {Fabrycky}}{{Perets} \&
  {Fabrycky}}{2009}]{Perets2009}
{Perets} H.~B.,  {Fabrycky} D.~C.,  2009, \mn@doi [\apj]
  {10.1088/0004-637X/697/2/1048}, \href
  {https://ui.adsabs.harvard.edu/#abs/2009ApJ...697.1048P} {697, 1048}

\bibitem[\protect\citeauthoryear{{Pritchet} \& {Glaspey}}{{Pritchet} \&
  {Glaspey}}{1991}]{Pritchet1991}
{Pritchet} C.~J.,  {Glaspey} J.~W.,  1991, \mn@doi [\apj] {10.1086/170029},
  \href {http://adsabs.harvard.edu/abs/1991ApJ...373..105P} {373, 105}

\bibitem[\protect\citeauthoryear{{Pr{\v{s}}a} \& {Zwitter}}{{Pr{\v{s}}a} \&
  {Zwitter}}{2005}]{2005ApJ...628..426P}
{Pr{\v{s}}a} A.,  {Zwitter} T.,  2005, \mn@doi [\apj] {10.1086/430591}, \href
  {https://ui.adsabs.harvard.edu/abs/2005ApJ...628..426P} {628, 426}

\bibitem[\protect\citeauthoryear{{Raddi} et~al.,}{{Raddi}
  et~al.}{2017}]{2017MNRAS.472.4173R}
{Raddi} R.,  et~al., 2017, \mn@doi [\mnras] {10.1093/mnras/stx2243}, \href
  {https://ui.adsabs.harvard.edu/abs/2017MNRAS.472.4173R} {472, 4173}

\bibitem[\protect\citeauthoryear{{Ricker} et~al.,}{{Ricker}
  et~al.}{2015}]{2015JATIS...1a4003R}
{Ricker} G.~R.,  et~al., 2015, \mn@doi [Journal of Astronomical Telescopes,
  Instruments, and Systems] {10.1117/1.JATIS.1.1.014003}, \href
  {https://ui.adsabs.harvard.edu/abs/2015JATIS...1a4003R} {1, 014003}

\bibitem[\protect\citeauthoryear{{Rodr{\'\i}guez}, {L{\'o}pez-Gonz{\'a}lez}  \&
  {L{\'o}pez de Coca}}{{Rodr{\'\i}guez} et~al.}{2000}]{2000A&AS..144..469R}
{Rodr{\'\i}guez} E.,  {L{\'o}pez-Gonz{\'a}lez} M.~J.,   {L{\'o}pez de Coca} P.,
   2000, \mn@doi [\aaps] {10.1051/aas:2000221}, \href
  {https://ui.adsabs.harvard.edu/abs/2000A&AS..144..469R} {144, 469}

\bibitem[\protect\citeauthoryear{{Rucinski}}{{Rucinski}}{1993}]{1993PASP..105.1433R}
{Rucinski} S.~M.,  1993, \mn@doi [\pasp] {10.1086/133326}, \href
  {https://ui.adsabs.harvard.edu/abs/1993PASP..105.1433R} {105, 1433}

\bibitem[\protect\citeauthoryear{{Sahu} et~al.,}{{Sahu}
  et~al.}{2019}]{sahu_2019}
{Sahu} S.,  et~al., 2019, \mn@doi [\apj] {10.3847/1538-4357/ab11d0}, \href
  {https://ui.adsabs.harvard.edu/abs/2019ApJ...876...34S} {876, 34}

\bibitem[\protect\citeauthoryear{{Salvatier}, {Wiecki}  \&
  {Fonnesbeck}}{{Salvatier} et~al.}{2015}]{2015arXiv150708050S}
{Salvatier} J.,  {Wiecki} T.,   {Fonnesbeck} C.,  2015, arXiv e-prints, \href
  {https://ui.adsabs.harvard.edu/abs/2015arXiv150708050S} {p. arXiv:1507.08050}

\bibitem[\protect\citeauthoryear{{Sandquist} \& {Shetrone}}{{Sandquist} \&
  {Shetrone}}{2003}]{Sandquist2003}
{Sandquist} E.~L.,  {Shetrone} M.~D.,  2003, \mn@doi [\aj] {10.1086/368139},
  \href {http://adsabs.harvard.edu/abs/2003AJ....125.2173S} {125, 2173}

\bibitem[\protect\citeauthoryear{{Selam}}{{Selam}}{2004}]{2004A&A...416.1097S}
{Selam} S.~O.,  2004, \mn@doi [\aap] {10.1051/0004-6361:20034578}, \href
  {https://ui.adsabs.harvard.edu/abs/2004A&A...416.1097S} {416, 1097}

\bibitem[\protect\citeauthoryear{{Sindhu}, {Subramaniam}  \& {Radha}}{{Sindhu}
  et~al.}{2018}]{Sindhu2018}
{Sindhu} N.,  {Subramaniam} A.,   {Radha} C.~A.,  2018, \mn@doi [\mnras]
  {10.1093/mnras/sty2283}, \href
  {https://ui.adsabs.harvard.edu/#abs/2018MNRAS.481..226S} {481, 226}

\bibitem[\protect\citeauthoryear{{Sindhu} et~al.,}{{Sindhu}
  et~al.}{2019}]{Sindhu2019}
{Sindhu} N.,  et~al., 2019, \mn@doi [\apj] {10.3847/1538-4357/ab31a8}, \href
  {https://ui.adsabs.harvard.edu/abs/2019ApJ...882...43S} {882, 43}

\bibitem[\protect\citeauthoryear{{Sindhu} et~al.,}{{Sindhu}
  et~al.}{2020}]{Sindhu2020}
{Sindhu} N.,  et~al., 2020, {Detection of White Dwarf companions to Blue
  Straggler Stars from UVIT observations of M67} (\mn@eprint {arXiv}
  {1908.01573}), \mn@doi{10.1017/S1743921319006975}

\bibitem[\protect\citeauthoryear{{Subramaniam}, {Pandey}, {Jadhav}  \&
  {Sahu}}{{Subramaniam} et~al.}{2020}]{Subramaniam2020}
{Subramaniam} A.,  {Pandey} S.,  {Jadhav} V.~V.,   {Sahu} S.,  2020, \mn@doi
  [\japa] {10.1007/s12036-020-09668-1}, \href
  {https://ui.adsabs.harvard.edu/abs/2020JApA...41...45S} {41, 45}

\bibitem[\protect\citeauthoryear{{Theano Development Team}}{{Theano Development
  Team}}{2016}]{exoplanet:theano}
{Theano Development Team} 2016, arXiv e-prints, abs/1605.02688

\bibitem[\protect\citeauthoryear{{Tkachenko} et~al.,}{{Tkachenko}
  et~al.}{2013}]{2013A&A...556A..52T}
{Tkachenko} A.,  et~al., 2013, \mn@doi [\aap] {10.1051/0004-6361/201220978},
  \href {https://ui.adsabs.harvard.edu/abs/2013A&A...556A..52T} {556, A52}

\bibitem[\protect\citeauthoryear{{Vanderburg} \& {Johnson}}{{Vanderburg} \&
  {Johnson}}{2014}]{2014PASP..126..948V}
{Vanderburg} A.,  {Johnson} J.~A.,  2014, \mn@doi [\pasp] {10.1086/678764},
  \href {https://ui.adsabs.harvard.edu/abs/2014PASP..126..948V} {126, 948}

\bibitem[\protect\citeauthoryear{{Wenger} et~al.,}{{Wenger}
  et~al.}{2000}]{2000A&AS..143....9W}
{Wenger} M.,  et~al., 2000, \mn@doi [\aaps] {10.1051/aas:2000332}, \href
  {https://ui.adsabs.harvard.edu/abs/2000A&AS..143....9W} {143, 9}

\bibitem[\protect\citeauthoryear{{Wilson} \& {Devinney}}{{Wilson} \&
  {Devinney}}{1971}]{1971ApJ...166..605W}
{Wilson} R.~E.,  {Devinney} E.~J.,  1971, \mn@doi [\apj] {10.1086/150986},
  \href {https://ui.adsabs.harvard.edu/abs/1971ApJ...166..605W} {166, 605}

\bibitem[\protect\citeauthoryear{{Yakut} et~al.,}{{Yakut}
  et~al.}{2009}]{2009A&A...503..165Y}
{Yakut} K.,  et~al., 2009, \mn@doi [\aap] {10.1051/0004-6361/200911918}, \href
  {https://ui.adsabs.harvard.edu/abs/2009A&A...503..165Y} {503, 165}

\bibitem[\protect\citeauthoryear{{van den Berg}, {Tagliaferri}, {Belloni}  \&
  {Verbunt}}{{van den Berg} et~al.}{2004}]{VandenBerg2004Chandra}
{van den Berg} M.,  {Tagliaferri} G.,  {Belloni} T.,   {Verbunt} F.,  2004,
  \mn@doi [\aap] {10.1051/0004-6361:20031642}, \href
  {http://adsabs.harvard.edu/abs/2004A26A...418..509V} {418, 509}

\makeatother
\end{thebibliography}

\appendix

\section{Supplementary figures}

\begin{figure*}
    \centering
    \includegraphics[width=0.99\textwidth ]{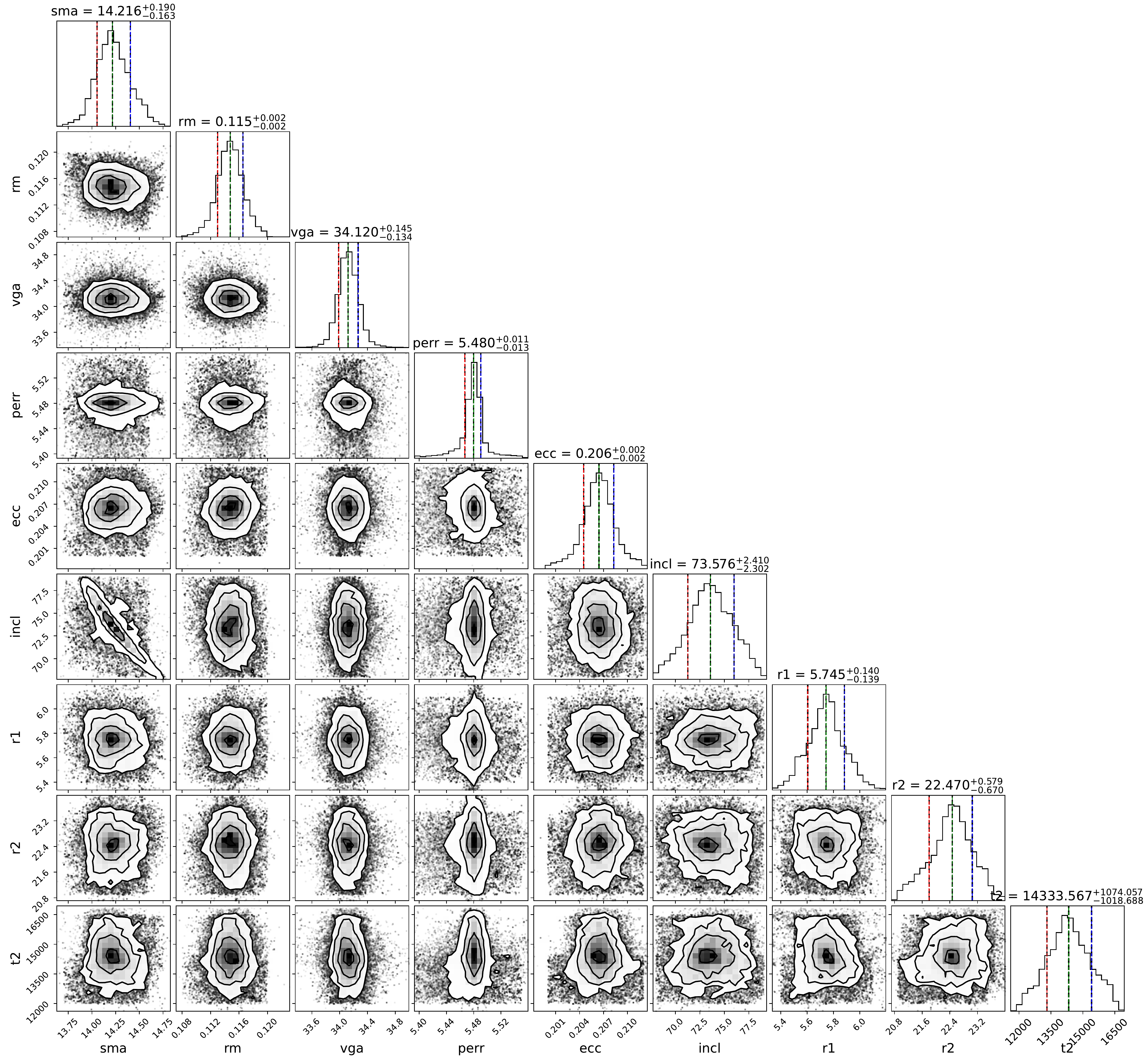}
   \caption{Corner plot obtained from the MCMC sampler for the stellar and orbital parameters of WOCS 1007. The red and blue lines represent the 16th ($-1\sigma$) and 84th ($+1\sigma$) percentile of the distribution, respectively. }
   \label{fig:mcmc}
\end{figure*}

\begin{figure*}
\centering
    \begin{subfigure}{.9\textwidth}
    \centering
    \includegraphics[width=\linewidth]{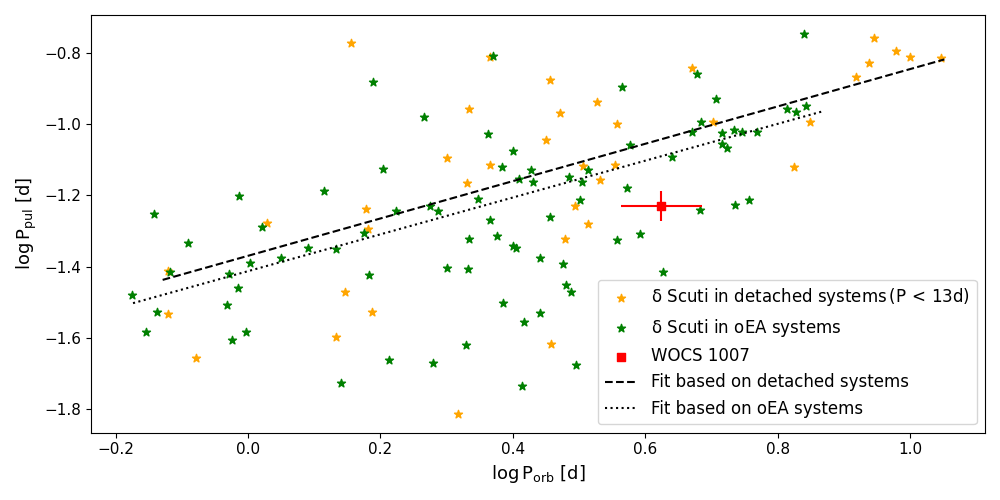}
    \caption{}\label{fig:pvp}
    \end{subfigure} %

    \hfill
    
    \begin{subfigure}{.9\textwidth}
    \centering
    \includegraphics[width=\linewidth]{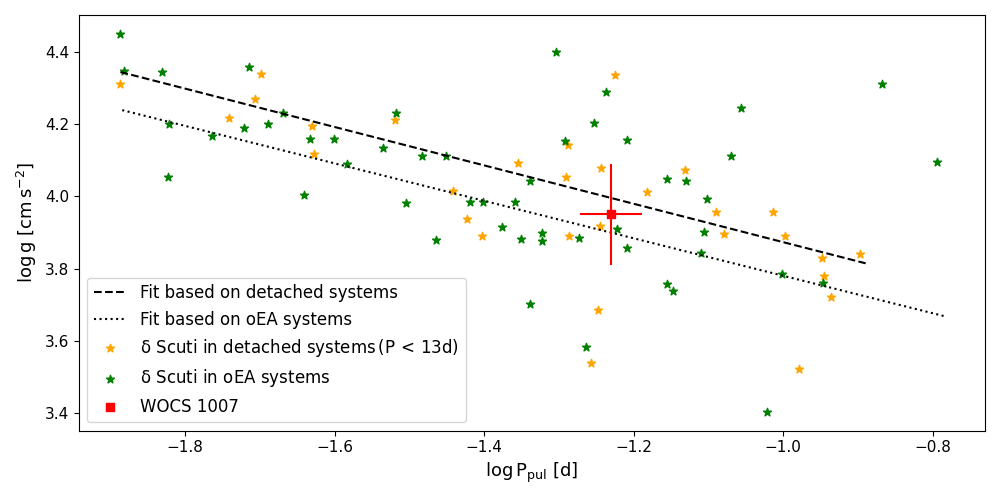}
    \caption{}\label{fig:pvg} 
    \end{subfigure} %
   
\RawCaption{\caption{Locations of the WOCS 1007 primary component (red) and other $\delta$ Scuti stars in the P$_{\rm orb}$--P$_{\rm pul}$ (top panel) and P$_{\rm pul}$--log g (bottom panel) diagram. The yellow and green points represent the $\delta$ Scuti stars in detached \citep{2020AcA....70..265L} and semi-detached or oEA systems (oscillating eclipsing systems of Algol type; \citealt{2017MNRAS.465.1181L}), respectively. The dashed and dotted lines represent the empirical linear fits obtained for detached systems by \citet{2020AcA....70..265L} and oEA systems by \citet{2017MNRAS.465.1181L}, respectively.}
\label{fig:relations}}
\end{figure*}

\bsp	
\label{lastpage}
\end{document}